\documentclass[prb,aps,showpacs,twocolumn,longbibliography]{revtex4-1}
\usepackage{times}
\usepackage{graphicx}
\usepackage{float}
\usepackage{latexsym,amsmath,amssymb,bm,euscript}
\usepackage{color}
\usepackage{epstopdf}
\usepackage[colorlinks=true,linkcolor=blue,citecolor=blue,urlcolor=blue]{hyperref}
\usepackage{hyperref}

\begin{document}

\title{Chiral spin liquid with spinon Fermi surfaces in spin-$1/2$ triangular Heisenberg model}
\author{Shou-Shu Gong$^{1}$, Wayne Zheng$^2$, Mac Lee$^{3,4}$, Yuan-Ming Lu$^2$, D. N. Sheng$^3$}
\affiliation{$^1$Department of Physics, Beihang University, Beijing 100191, China\\
$^2$Department of Physics, Ohio State University, Columbus, Ohio 43210, USA\\
$^3$Department of Physics and Astronomy, California State University, Northridge, CA 91330, USA\\
$^4$Department of Physics, University of California, San Diego, CA 92093, USA}

\begin{abstract}
We study the interplay of competing interactions in spin-$1/2$ triangular Heisenberg model through tuning the first- ($J_1$), second- ($J_2$), and third-neighbor ($J_3$) couplings.
Based on large-scale density matrix renormalization group calculation, we identify a quantum phase diagram of the system and discover a new {\it gapless} chiral spin liquid (CSL) phase in the intermediate $J_2$ and $J_3$ regime.
This CSL state spontaneously breaks time-reversal symmetry with finite scalar chiral order, and it has gapless excitations implied by a vanishing spin triplet gap and a finite central charge on the cylinder. 
Moreover, the central charge grows rapidly with the cylinder circumference, indicating emergent spinon Fermi surfaces. 
To understand the numerical results we propose a parton mean-field spin liquid state, the $U(1)$ staggered flux state, which breaks time-reversal symmetry with chiral edge modes by adding a Chern insulator mass to Dirac spinons in the $U(1)$ Dirac spin liquid. 
This state also breaks lattice rotational symmetries and possesses two spinon Fermi surfaces driven by nonzero $J_2$ and $J_3$, which naturally explains the numerical results. 
To our knowledge, this is the first example of a gapless CSL state with coexisting spinon Fermi surfaces and chiral edge states, demonstrating the rich family of novel phases emergent from competing interactions in triangular-lattice magnets.
\end{abstract}

\maketitle

{\it Introduction.} Quantum spin liquids (QSLs) are novel quantum phases of matter, which do not exhibit any symmetry-breaking orders even at zero temperature~\cite{balents2010, savary2016, Zhou2017} but feature long-range entanglement and fractionalized excitations~\cite{wen1991, senthil2000, senthil2001, Wen2013}.
QSLs have been studied extensively in the past few decades, due to their important role in understanding strongly correlated materials and potential application in topological quantum computation~\cite{anderson1973, Read1991, sachdev1992, lee2006, kitaev2006}.
While gapped QSLs have been classified and characterized systematically, there is much less understanding on gapless QSLs and how they could be realized in materials. 
Although a gapless QSL with Dirac cones of spinons has been shown to exist in the exactly soluble Kitaev model~\cite{kitaev2006}, so far there is no definitive evidence that a Dirac spin liquid has been realized in any magnetic materials~\cite{savary2016, hermanns2018}.
A more exotic state is the gapless QSL with spinon Fermi surfaces (SFSs)~\cite{ioffe1989,nagaosa1990, lee1992}.
Such a QSL has an extensive number of low-energy excitations, and was shown to be stabilized by four-spin ring-exchange couplings that can arise from strong charge fluctuations in weak Mott insulators~\cite{misguich1999, lesik2005, sheng2009, block2011, he2018}.

Experimentally, many QSL candidate materials fall into the family of layered spin-$1/2$ magnets on the triangular lattice, such as the organic salts~\cite{shimizu2003, kurosaki2005, yamashita2008, yamashita2009, yamashita2010} and the transition metal dichalcogenides~\cite{law2017, yu2017, ribak2017, klanjsek2017}.
Specific heat and thermal transport measurements point towards the presence of extensive mobile gapless spin excitations, which appear to be consistent with a gapless QSL with SFSs~\cite{yamashita2008, yamashita2010, yu2017}.
These materials are considered to be weak Mott insulators with strong charge fluctuations, which may induce such gapless QSL behaviors~\cite{misguich1999, lesik2005, sheng2009, block2011, he2018}.
However, a direct study on the triangular Hubbard model suggests a possible gapped chiral spin liquid (CSL) phase in the intermediate $U$ region~\cite{szasz2018}.
Therefore, a clear theoretical understanding on the mechanism to realize gapless QSLs in these layered quasi-two-dimensional magnets is still lacking. 

Another route to QSL is through competing interactions between different neighboring sites, such as the kagome compound kapellasite~\cite{fak2012} and $J_1$-$J_2$-$J_3$ kagome model~\cite{gong2014kagome, gong2015}. 
Recently, competing interactions have also been found essential to understand possible QSLs in the triangular-lattice rare-earth compounds~\cite{paddison2016, zhu2018, zhang2018, li2015, shen2018} and delafossite oxides~\cite{kimura2006, ye2007, kadowaki1990, seki2008}.
Indeed, a QSL phase has been found in the spin-$1/2$ $J_1$-$J_2$ triangular Heisenberg model (THM) although its nature has not been established~\cite{kaneko2014, iqbal2016, campbell2015, zhuzhenyue2015, Hu2015,Zheng2015,Lu2016, mcculloch2016}.
Therefore, understanding how QSL phases emerge from competing interactions is an important issue in order to discover new QSL materials~\cite{ivanov1995, yao2018, iaconis2018}. 

\begin{figure}[t]
\includegraphics[width = 0.8\linewidth]{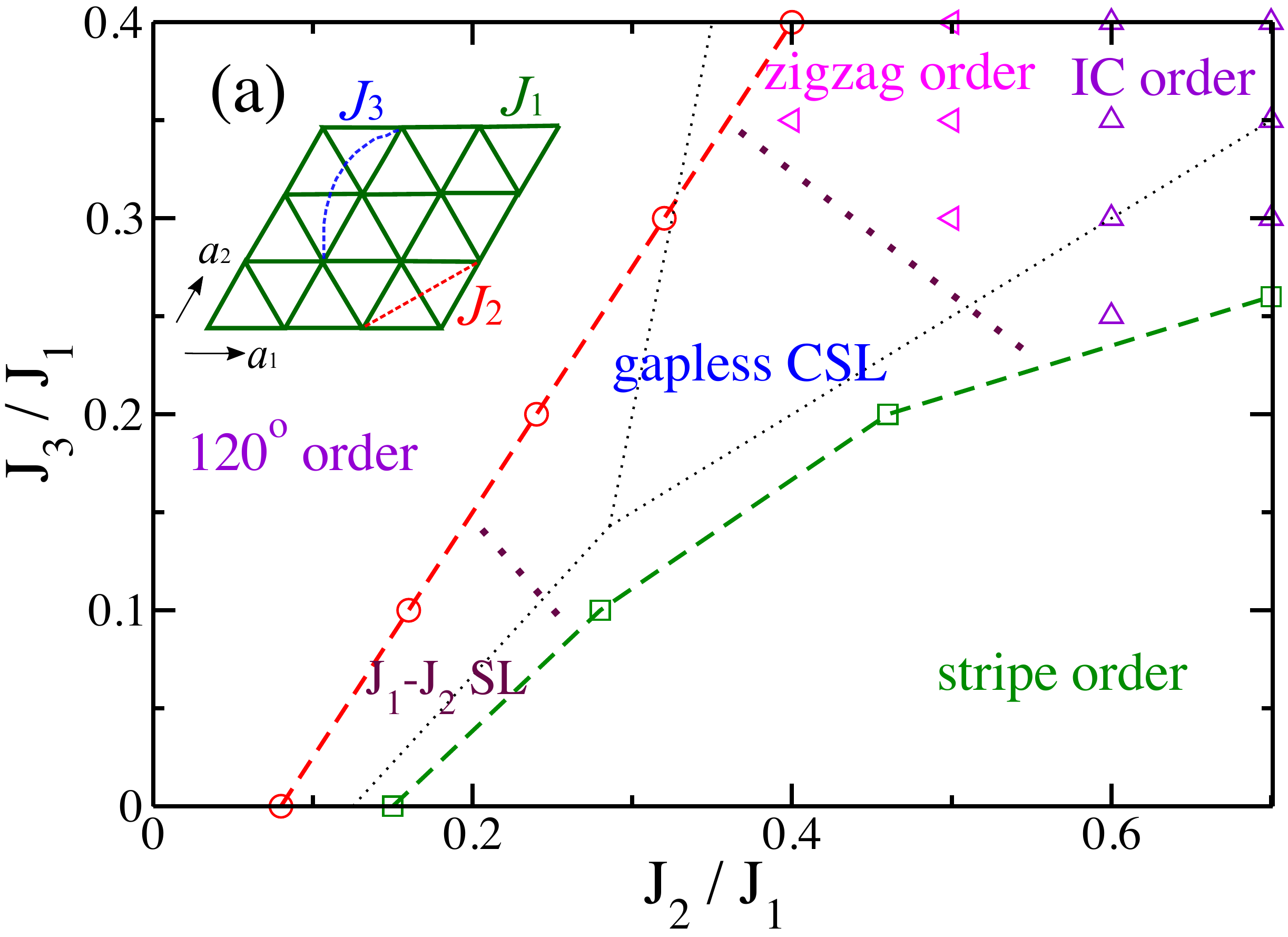}
\includegraphics[width = 0.325\linewidth]{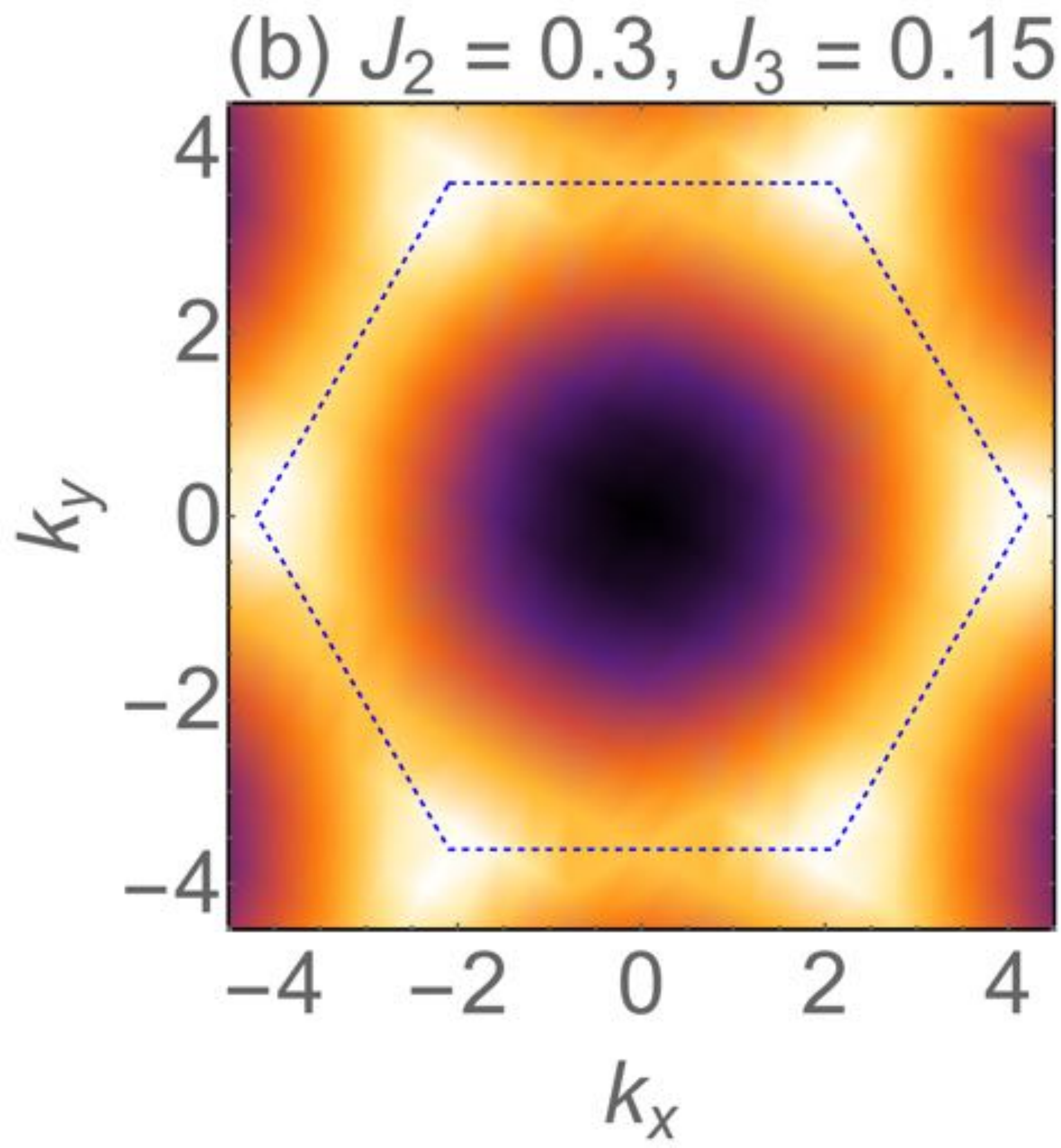}
\includegraphics[width = 0.325\linewidth]{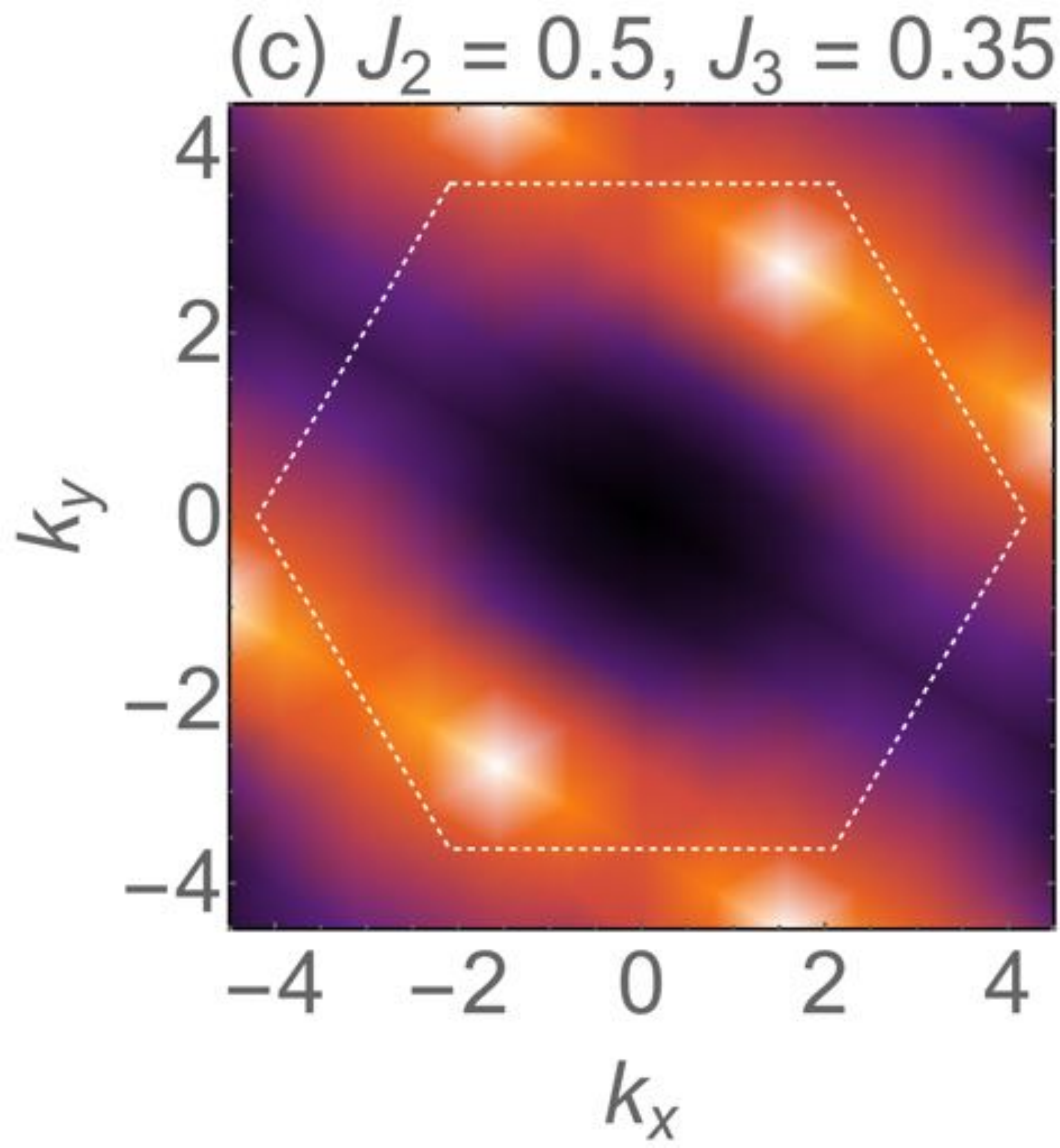}
\includegraphics[width = 0.325\linewidth]{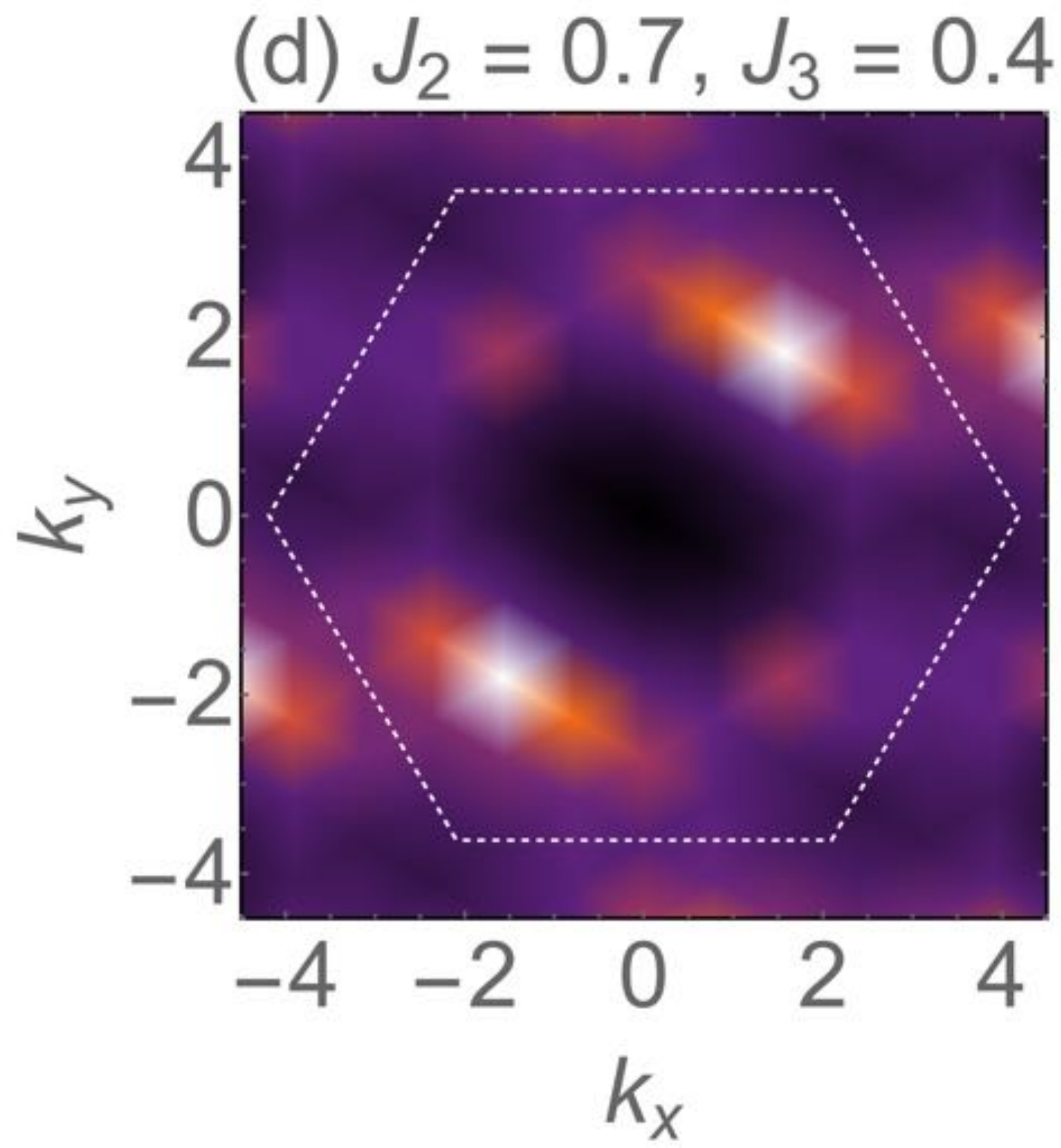}
\caption{Quantum phase diagram of spin-$1/2$ $J_1$-$J_2$-$J_3$ THM.
The inset shows the model on the YC geometry.
We find a $120^0$ order, a stripe order, a zigzag order, an incommensurate (IC) order, and a gapless chiral spin liquid (CSL) phase in the neighbor of the previously found $J_1$-$J_2$ spin liquid (SL) phase.
The colored dotted-lines are schematic phase boundaries, and the black dotted-lines are the classical phase boundaries of the $120^0$, stripe, and zigzag orders. Static spin structure factors of the gapless CSL (b), the zigzag state (c), and the incommensurate state (d) on the YC8 cylinder.}
\label{fig:phase}
\end{figure}

In this Rapid communication, we systematically study the spin-$1/2$ $J_1$-$J_2$-$J_3$ THM using density-matrix renormalization group (DMRG) method and the parton construction. The model Hamiltonian is given as
\begin{equation}
 H = J_1 \sum_{\langle i,j \rangle} {\bf S}_i \cdot {\bf S}_j
 + J_2 \sum_{\langle\langle i,j \rangle\rangle} {\bf S}_i \cdot {\bf S}_j
  + J_3 \sum_{\langle\langle\langle i,j \rangle\rangle\rangle} {\bf S}_i \cdot {\bf S}_j,
\label{eq:ham}
\end{equation}
where $J_1, J_2, J_3$ are the first-, second-, and third NN interactions as shown in the inset of Fig.~\ref{fig:phase}(a).
We choose $J_1 = 1.0$ as the energy scale.
In the coupling range $0 \leq J_2/J_1 \leq 0.7, 0 \leq J_3/J_1 \leq 0.4$, besides the previously found $J_1$-$J_2$ spin liquid and different magnetic orders, we identify a new gapless CSL phase as shown in Fig.~\ref{fig:phase}(a).
This CSL state spontaneously breaks time-reversal symmetry (TRS) with a finite scalar chiral order. We also observe spin pumping upon flux insertion, similar to the charge pumping in Laughlin-type fractional quantum Hall states, indicative of a chiral edge mode, which is further confirmed by the entanglement spectrum. Finite-size scaling of spin triplet gap on the square-like clusters shows a vanished spin gap. The gapless nature is further supported by the bipartite entanglement entropy, which exhibits a logarithmic correction of the area law versus subsystem length, leading to a finite central charge.
The central charge grows with the cylinder circumference consistent with a QSL with  emergent SFSs.
We propose a staggered flux state in the Abrikosov-fermion representation of spin-$1/2$ operators, which explains the coexistence of chiral edge mode and SFSs observed in this gapless CSL.

We study the system by using DMRG with $SU(2)$ symmetry~\cite{white1992, mcculloch2002}.
We use cylinder geometry with periodic boundary conditions along circumference direction and open boundary conditions along extended direction.
The lattice vectors are defined as ${\bf a}_1 = (1, 0)$ and ${\bf a}_2 = ( \frac{1}{2}, \frac{\sqrt{3}}{2} )$.
Two geometries named YC and XC cylinders are studied, both having extended direction along ${\bf a}_1$.
For the YC and XC cylinders, circumference direction is along ${\bf a}_2$ and perpendicular to ${\bf a}_1$, respectively.
The cylinders are denoted as YC$L_y - L_x$ and XC$L_y - L_x$ with $L_y$ and $L_x$ being the numbers of sites along circumference and extended directions.
We study the systems with $L_y = 5 - 12$ by keeping up to $8000$ $SU(2)$ states (equivalent to about 24000 $U(1)$ states) to obtain accurate results with truncation error less than $10^{-5}$ in most calculations.

{\it Quantum phase diagram.} We demonstrate the quantum phase diagram in Fig.~\ref{fig:phase}(a).
With growing $J_2$ and $J_3$, we find different magnetically ordered phases and QSL phases.
In Fig.~\ref{fig:phase}(a), the black dotted lines denote the classical phase boundaries of the 120$^0$, stripe, and zigzag orders.
We also find an incommensurate (IC) magnetic order in the neighbor of the zigzag order, consistent with previous spin-wave calculations~\cite{ivanov1995}.
The incommensurate order might be considered as the zigzag order with an incommensurate modulation (see Supplemental Material~\cite{suppl}).
In the presence of quantum fluctuations, we find a new gapless CSL phase near the triple point of the classical orders, which sits at the neighbor of the $J_1$-$J_2$ SL.
By computing spin and dimer correlation functions, we find featureless spin and dimer structure factors that indicate the absence of spin and dimer orders in the CSL state~\cite{suppl}.
Next, we further characterize the nature of this new CSL state.

\begin{figure}[t]
\includegraphics[width = 1\linewidth]{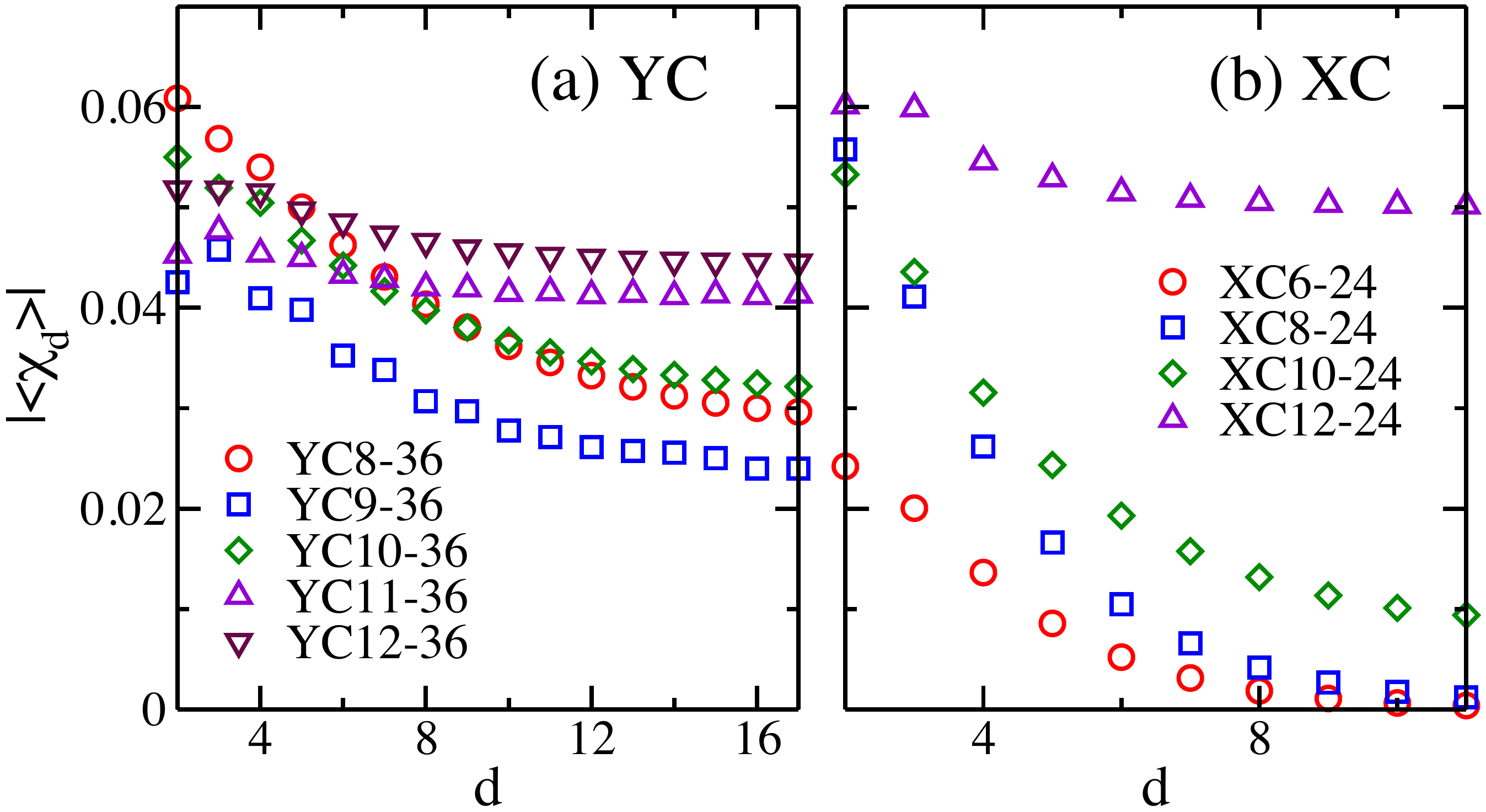}
\caption{Finite scalar chiral order of the CSL state at $J_2 = 0.3, J_3 = 0.15$.
(a) and (b) are the scalar chiral order measured from the boundary to the bulk on the YC and XC cylinders. The scalar chiral order $\langle \chi \rangle = \langle {\bf S}_1 \cdot ({\bf S}_2 \times {\bf S}_3) \rangle$ is defined for the three spins ${\bf S}_i$ $(i = 1, 2, 3)$ for each triangle, and $d$ is the distance of the triangle from the edge. The chiral orders of all the triangles have the same chiral direction.
}
\label{fig:chiral}
\end{figure}

{\it Spontaneous time-reversal symmetry breaking.} To detect spontaneous TRS breaking, we use complex-valued wavefunction, which has been applied in DMRG to find chiral ground states in different systems~\cite{gong2014kagome, zhu2016}.
If TRS is spontaneously broken, the system is featured by finite scalar chiral order $\langle \chi \rangle = \langle {\bf S}_{1} \cdot ({\bf S}_{2} \times {\bf S}_{3}) \rangle$, where ${\bf S}_{i}$ $(i = 1, 2, 3)$ label the three spins on each triangle.
On the YC cylinder with both even and odd $L_y$, we find a nonzero chiral order in the bulk of cylinder with a large circumference, as shown in Fig.~\ref{fig:chiral}(a) for $J_2 = 0.3, J_3 = 0.15$.
In these states, the chiral orders of all the up- and down-triangles have the same sign, and the chiral order grows more robust as the circumference increases.
On the XC cylinder shown in Fig.~\ref{fig:chiral}(b), the chiral order vanishes in the bulk for small circumference but becomes stable on the wide XC12 cylinder.
Combining these results we conclude a CSL state with spontaneous TRS breaking.

\begin{figure}[t]
\includegraphics[width = 0.48\linewidth]{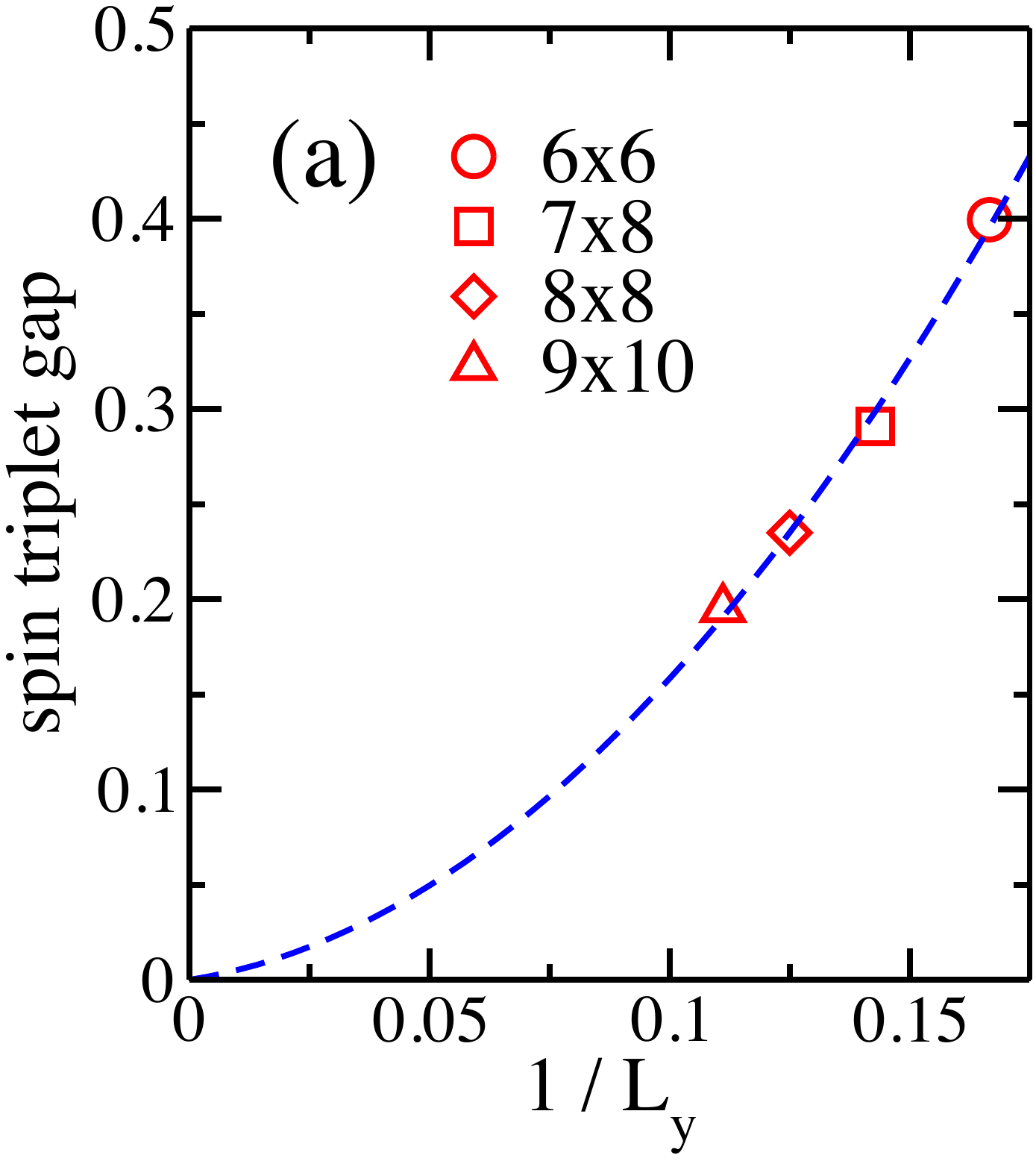}
\includegraphics[width = 0.495\linewidth]{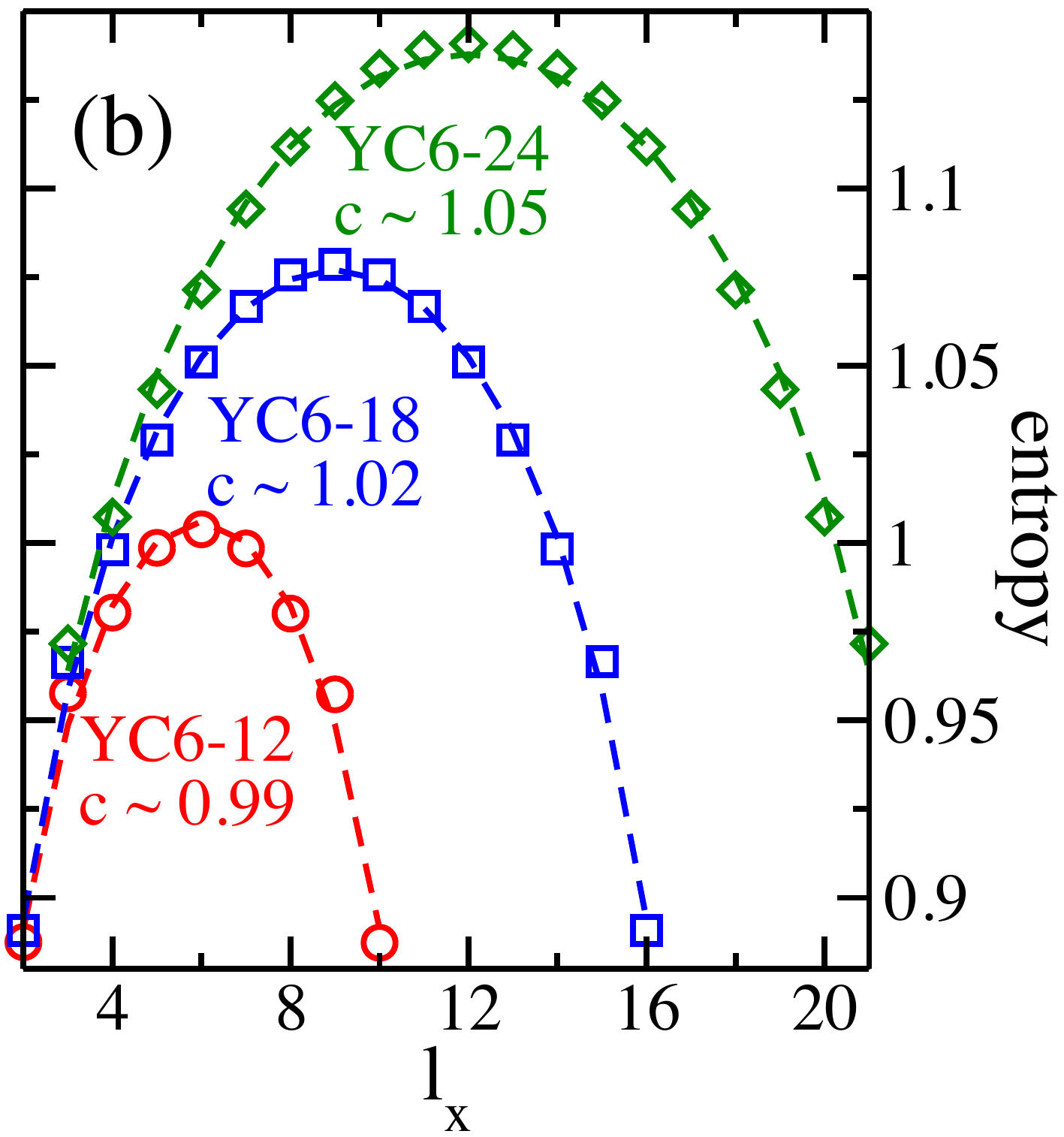}
\includegraphics[width = 0.495\linewidth]{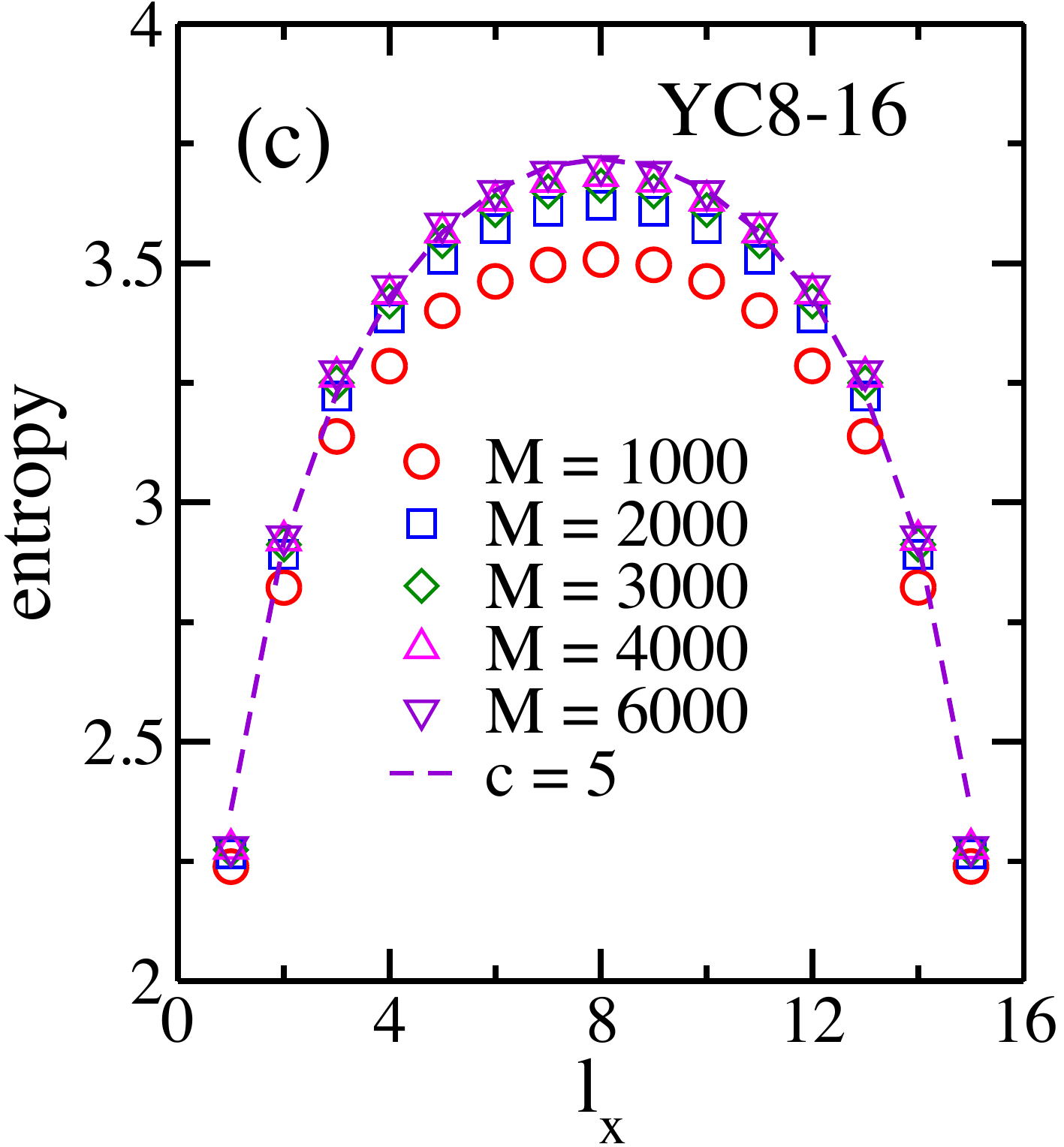}
\includegraphics[width = 0.475\linewidth]{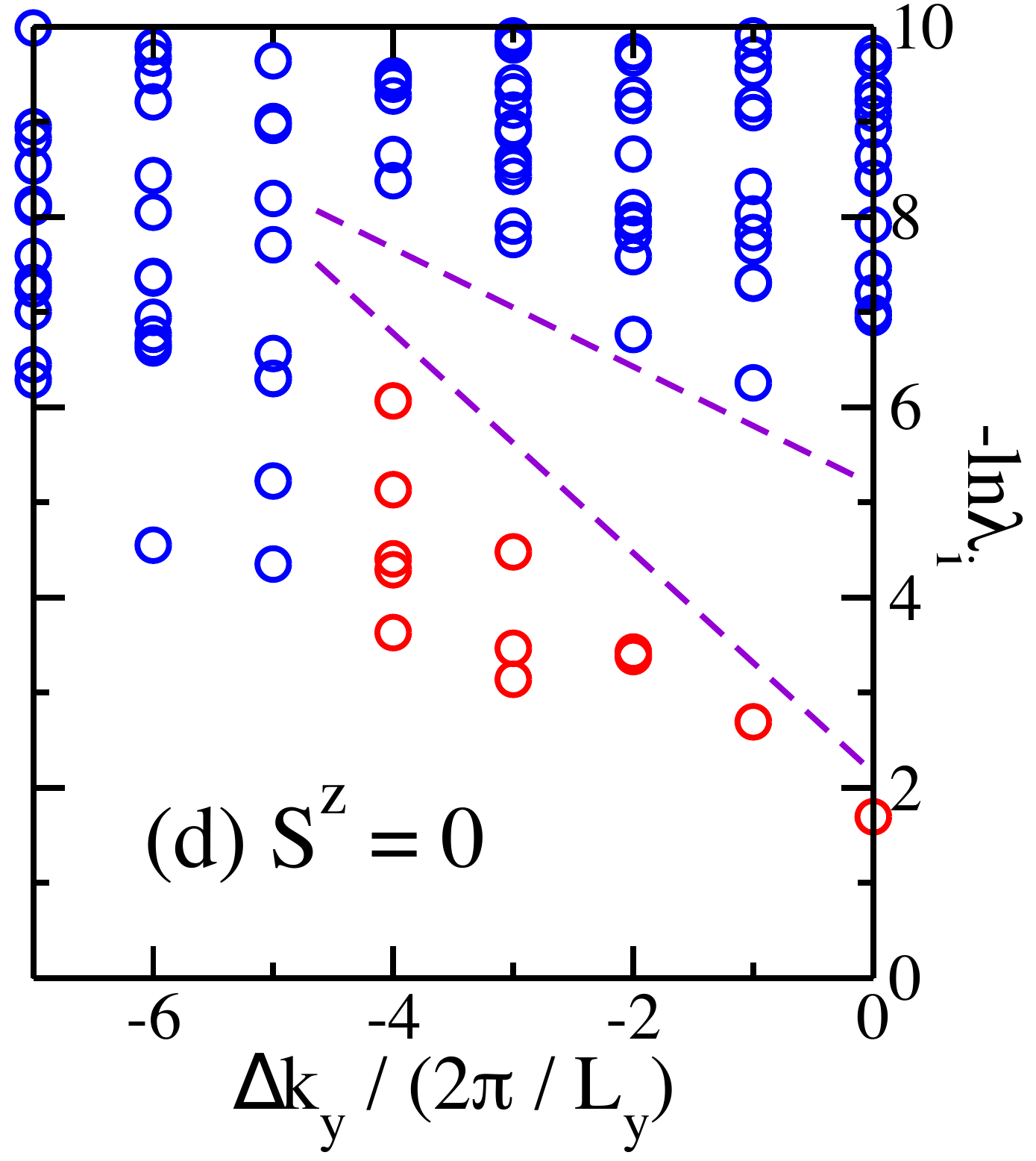}
\caption{(a) Size scaling of the spin triplet gap obtained on the square-like clusters.
(b) Entanglement entropy versus subsystem length $l_x$ on the YC6 cylinders with different $L_x$.
(c) Entanglement entropy on the YC8-16 cylinder by keeping different $SU(2)$ state numbers.
The dash lines denote the fitting of entropy following the formula $S(l_x) = (c/6) \ln[ (L_x / \pi) \sin( l_x \pi / L_x ) ] + g$, giving central charge $c \simeq 1$ for the YC6 cylinder and $c \simeq 5$ for the YC8-16 cylinder.
(d) Entanglement spectrum labeled by the quantum numbers total spin $S^z = 0$ and relative momentum along the $y$ direction $\Delta k_y$. $\lambda_i$ is the eigenvalue of reduced density matrix. The red circles denote the near degenerate pattern $\{1, 1, 2, 3, 5\}$ of the low-lying spectrum.
}
\label{fig:gap}
\end{figure}

{\it Spin triplet gap and entanglement characterization.} We calculate the spin triplet gap by obtaining the ground state (in the $S = 0$ sector) on long cylinder and then sweeping the $S = 1$ sector for the middle $N_x$ columns~\cite{Yan2011}, which gives the gap of the middle $N_x \times L_y$ system.
We find that the gap versus $1/N_x$ shows length dependence~\cite{suppl}.
To estimate the gap in the 2D limit and avoid 1D physics, we extrapolate the gap data of the square-like clusters as shown in Fig.~\ref{fig:gap}(a). The gap drops fast as a function of $1/L_y$ and smoothly scales to zero, suggesting gapless spin-triplet excitations.

Furthermore, we study entanglement entropy versus subsystem length $l_x$ by cutting the cylinder into two parts.
Since the real-valued wavefunction is a superposition of the two chiral states with opposite chiralities, it has a higher entanglement entropy and is harder to converge to; thus we also use complex-valued wavefunction to compute entanglement entropy.
As shown in Fig.~\ref{fig:gap}(b) for $J_2 = 0.3, J_3 = 0.15$ on the YC6 cylinder, the entropy shows a logarithmic correction of the area law and follows the behavior $S(l_x) = (c/6) \ln[ (L_x / \pi) \sin(l_x \pi / L_x) ] + g$~\cite{calabrese2004}, where $S(l_x)$ is the bipartite entanglement entropy, $c$ is the central charge, and $g$ is a non-universal constant.
The YC6 cylinders with different $L_x$ give a consistent central charge $c \simeq 1$.
For the YC8 cylinder, we choose $L_x = 16$ (the entropy for larger $L_x$ is much harder to converge and we show the results for $L_x = 24$ in Supplemental Material~\cite{suppl}, which are consistent with the fitted central charge $c=5$).
As shown in Fig.~\ref{fig:gap}(c), the entropy continues to grow with kept state number and converges very well by keeping $6000$ $SU(2)$ states, giving a large central charge of $c \simeq 5$.
The finite central charge supports the gapless nature of the CSL state. Once a 2D quantum state is confined to a 1D cylinder, the finite circumference quantizes the momentum around the cylinder.
The central charge of the 1D cylinder needs to sum over contributions from all quantized momenta.
Take the $U(1)$ Dirac spin liquid for example, the cylinder central charge $c = 2 - 1 = 1$ if the quantized momenta for each spin species only cross one Dirac cone, where the extra $-1$ accounts for the $U(1)$ gauge field fluctuations which gaps out the total spinon density fluctuation~\cite{Geraedts2016,suppl}.
Similarly $c \leq 3$ if the quantized momenta cross two Dirac cones (for each spin species), which is an upper bound for the central charge on a cylinder of any $L_y$.
The large central charge we found from DMRG is therefore inconsistent with the $U(1)$ Dirac spin liquid on triangular lattice~\cite{kaneko2014,iqbal2016,Lu2016}, but provides a strong evidence supporting emergent SFSs~\cite{ioffe1989,nagaosa1990,lesik2005,sheng2009}.
Now that each pair of crossings (one right mover and one left mover) between the quantized momenta and the SFSs contributes a unit of central charge, the total central charge of SFSs generally grows with $L_y$, with an upper bound of $c\leq N_w-1$ where $2N_w$ is the total number of crossings~\cite{suppl}.

For gapped CSL states, entanglement spectrum has a one-to-one correspondence with physical edge spectrum~\cite{li2008}.
Interestingly, for this gapless CSL state entanglement spectrum also shows a quasi-degenerate group of levels with the counting $\{ 1, 1, 2, 3, 5,\cdots\}$ agreeing with chiral $SU(2)_1$ conformal field theory~\cite{cft}, as shown in Fig.~\ref{fig:gap}(d).
This may be the first example of such novel states for interacting system, which demonstrates similar edge physics as the non-interacting $p+ip$ chiral superconductor with a gapless bulk spectrum~\cite{dubail2015, poilblanc2017}.

{\it The staggered flux state.} To understand the DMRG results, we propose a staggered flux state, whose mean-field ansatz is constructed in the Abrikosov-fermion representation of spin-$1/2$ operators~\cite{Abrikosov1965}
\begin{equation}
    \mathbf{S}_i = \frac{1}{4}\text{Tr}\left( \psi^{\dagger}_i \psi_i \bm{\sigma}^{T} \right), \quad
    \psi_i =
    \begin{pmatrix}
        f_{i, \uparrow} & f_{i, \downarrow} \\
        f_{i, \downarrow}^{\dagger} & -f_{i, \uparrow}^{\dagger}
    \end{pmatrix}
\end{equation}
The Heisenberg Hamiltonian $H=\sum_{\langle{ij}\rangle}J_{ij}\mathbf{S}_{i}\cdot\mathbf{S}_{j}$ is decoupled into the mean-field form as
\begin{equation}
    H_{\bf MF}=\frac{1}{8}\sum_{{ij}}J_{ij}\text{Tr}\left( \psi_{i}^{\dagger}u_{ij}\psi_{j}+h.c. \right)+\frac{1}{8}\sum_{{ij}}J_{ij}\text{Tr}\left( u_{ij}^{\dagger}u_{ij} \right) \nonumber
\end{equation}
with the mean-field amplitude $u_{ij} = \langle \psi_{i} \psi_{j}^{\dagger} \rangle=u_{ij}^\dagger$.
In the $U(1)$ QSL states all spinon pairing terms will vanish and thus
\begin{equation}\label{eq:}
    u_{ij}=
    \begin{pmatrix}
        -\bar{\chi}_{ij} & 0 \\
        0 & \chi_{ij}
    \end{pmatrix},
\end{equation}
where $\chi_{ij} = \sum_{\alpha}\langle f_{i, \alpha}^{\dagger}f_{j, \alpha} \rangle = \bar{\chi}_{ji}$. Then the mean-field ansatz can be simplified as
\begin{equation}
    H_{\bf MF} = \frac{J}{4}\sum_{\langle{ij}\rangle}\sum_{\alpha}\left(-\bar{\chi}_{ij}f_{i, \alpha}^{\dagger}f_{j, \alpha}+h.c.\right)+\frac{J}{4}\sum_{\langle{ij}\rangle}\left(|\chi_{ij}|^{2}\right) \nonumber
\end{equation}
where the mean-field ground state is at half-filling due to the single-occupancy constraint on the parton Hilbert space
\begin{equation}\label{single occupancy}
  \sum_{\alpha=\uparrow,\downarrow}f^\dagger_{i\alpha}f_{i\alpha}=1,~~~\forall~i.
\end{equation}
We consider a $U(1)$ spin liquid known as the staggered flux state~\cite{PhysRevB.65.165113,Li2017a,bieri2016}, where fermionic spinons transform under translations as follows
\begin{equation}
  f_{\mathbf{r},\alpha} \xrightarrow{T_{2}}(-)^{r_{1}}f_{\mathbf{r}+{\bf a}_2, \alpha}^{\dagger},~~f_{\mathbf{r},\alpha} \xrightarrow{T_{1}}f_{\mathbf{r}+{\bf a}_1,\alpha}.
    \label{eq:mag_trans}
\end{equation}
Although the mean-field ansatz doubles the unit cell (along ${\bf a}_2$ direction), the projected wavefunction preserves the lattice translation symmetries by ${\bf a}_{1,2}$.

\begin{figure}
    \centering
    \includegraphics[width=0.45\textwidth]{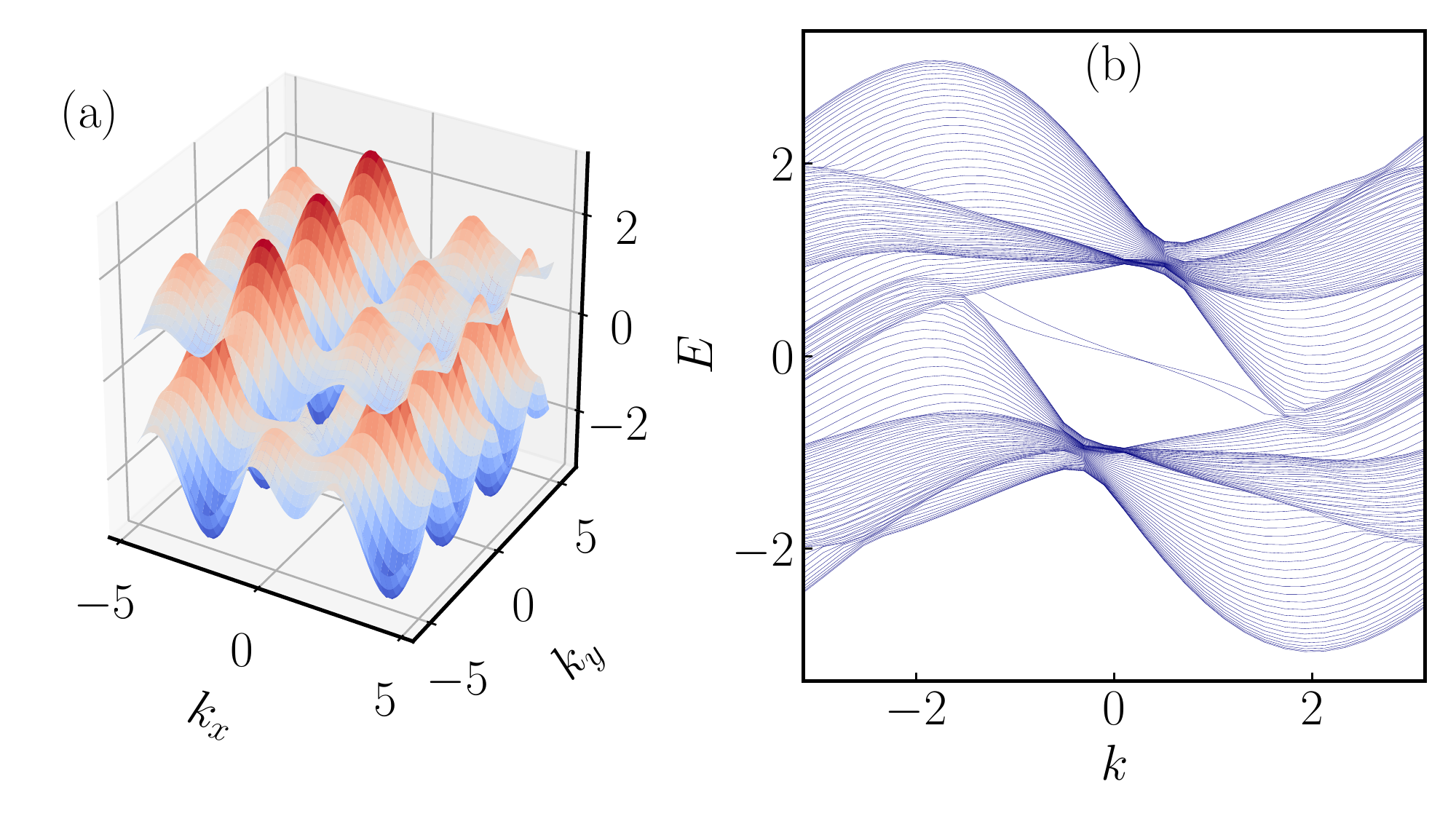}
    \includegraphics[width=0.45\textwidth]{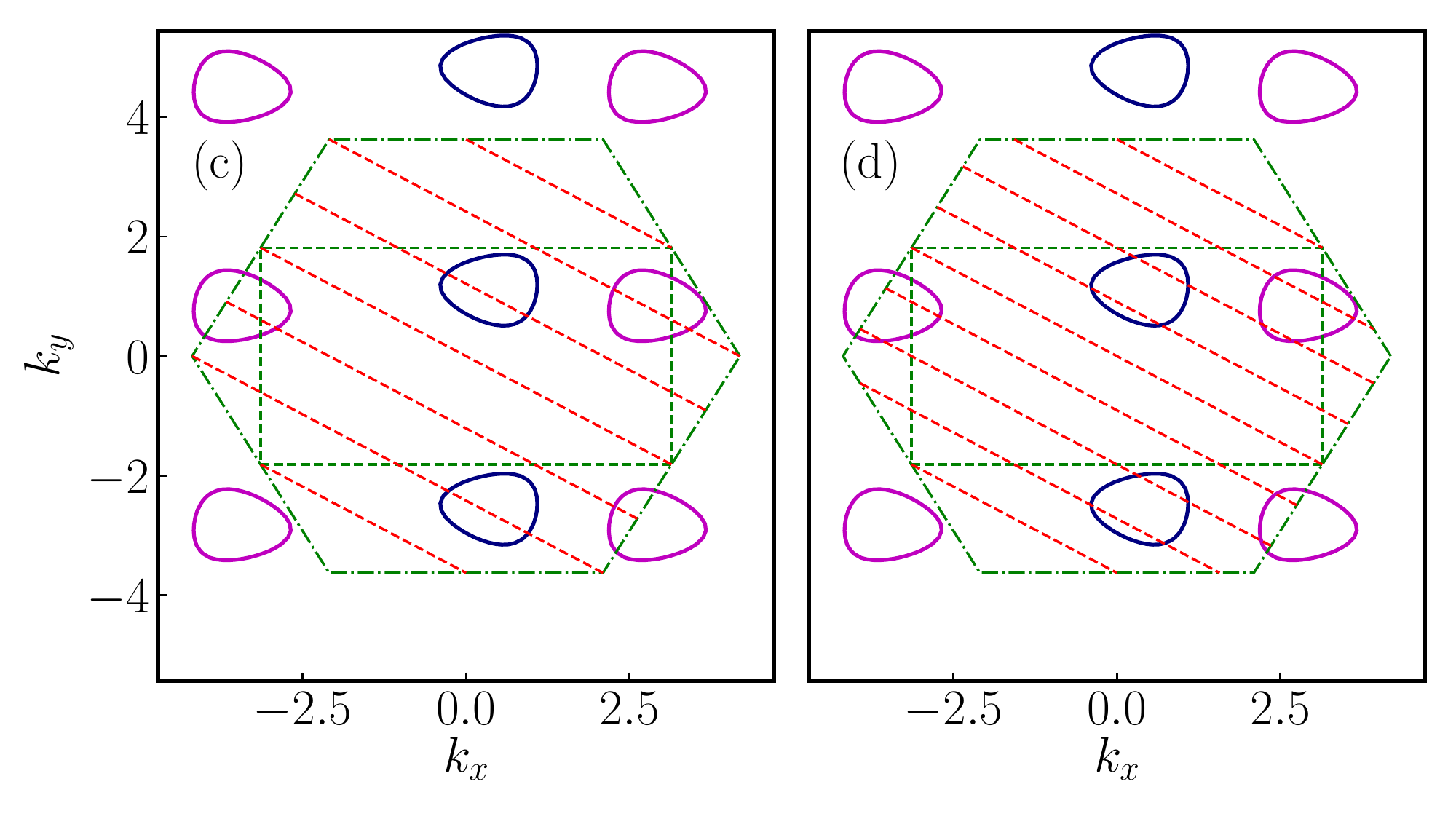}
    \caption{Mean-field ansatz of the staggered flux state with up to third NN mean-field amplitudes. (a) Spinon dispersion. (b) Edge spectrum on a cylinder geometry. (c) and (d) show how the 1d channels with quantized momentum $k_2=\frac{2\pi}{L_y}l_2,~l_2\in\mathbb{Z}$ cross the spinon Fermi surfaces (SFSs) on finite YC cylinders with $L_y=6, 8$. The dotted rectangle is the reduced Brillouin zone due to the doubling of unit cell in the mean-field ansatz. Pink and blue circles denote hole- and particle-like SFSs respectively.}
    \label{fig:stagger_mf}
\end{figure}

Since the spin model has couplings up to the 3rd NN sites, we consider the symmetry-allowed mean-field ansatz with hopping terms up to the 3rd NN, which are shown in Supplemental Material~\cite{suppl}.
The NN hopping ansatz reduces to the $\pi$-flux $U(1)$ QSL state~\cite{Lu2016} in the case of $\phi_1 = \phi_2 = \pi/2$ ($\phi_1, \phi_2$ are the phases of the NN hoppings), with a pair of Dirac spinons at half filling for each spin species.
The 2nd and 3rd NN hoppings can open up a direct gap at each Dirac cone, leading to a Chern number $C=\pm1$ of the lower spinon band and the chiral edge states shown in Fig.~\ref{fig:stagger_mf}(b). Meanwhile the 3rd NN hoppings can break the degeneracy of two Dirac cones, giving rise to one particle-like SFS around one Dirac point (blue in Figs.~\ref{fig:stagger_mf}(c-d)) and a hole-like SFS around the other Dirac point (pink in Figs.~\ref{fig:stagger_mf}(c-d)). Due to single-occupancy constraint Eq.~\eqref{single occupancy}, the particle-like SFS and hole-like SFS are perfectly compensated at half filling. Choosing mean-field parameters as $\chi=1.0, \phi_{1}=\phi_{2}=\pi/2, \lambda=1.0, \varphi_{1}=\varphi_{2}=\varphi_{3}=0, \rho=3.0, \gamma_{1}=\gamma_{2}=\gamma_{3}=\pi/2$~\cite{suppl}, the mean-field dispersion and edge spectrum of fermionic spinons are shown in Figs.~\ref{fig:stagger_mf}(a-b).

For further comparison with DMRG, we follow the YC cylinder geometry with quantized momentum $k_{2} = 2\pi l_{2}/L_{y}$ along $\mathbf{b}_{2}$ direction. In Fig.~\ref{fig:stagger_mf}(c) and (d) we depict how the 1D channels with quantized momenta $k_{2} = 2\pi l_{2}/L_{y}$ intersect with the two SFSs in the reduced Brillouin zone of the staggered flux state. On the YC6 cylinder, as shown in Fig. \ref{fig:stagger_mf}(c), there are $N_w = 2\times2=4$ pairs of gapless 1D modes crossing the SFSs (counting both spin species), constraining the central charge to be $c \leq N_w - 1 = 3$. On the YC8 cylinder, as shown in Fig.~\ref{fig:stagger_mf}(d), there are $N_w = 2\times4=8$ pairs of gapless 1D modes crossing the SFSs, restricting the central charge as $c \leq N_w - 1 = 7$. This is consistent with the observed $c \approx 1$ on YC6-24 cylinder (Fig.~\ref{fig:gap}(b)) and $c \approx5$ on YC8-16 cylinder (Fig.~\ref{fig:gap}(c)). Note that the number $N_w-1$ only bound the actual central charge from above, since symmetric backscatterings between these gapless 1D channels can further reduce the total central charge from $N_w - 1$~\cite{suppl}.

{\it Discussion.} The spin structure factor of the gapless CSL phase in Fig. \ref{fig:phase}(b) resembles that of the $U(1)$ Dirac spin liquid~\cite{iqbal2016}. Specifically, it exhibits high intensities on the edge and at the corner of the hexagonal Brillouin zone, which are associated with fermion bilinears and monopoles respectively in the $U(1)$ Dirac spin liquid~\cite{Song2018,Song2018a}. This suggests the proximity of the gapless CSL to the $U(1)$ Dirac state, which is indeed the case for the proposed staggered flux state.
We have also studied the phase transition from the $J_1$-$J_2$ SL phase to the gapless CSL phase.
The ground-state energy versus couplings is very smooth, suggesting a possible continuous phase transition~\cite{suppl}.
Interestingly, in the $J_1$-$J_2$ SL entanglement entropy also shows a logarithmic correction of the area law, which leads to a finite central charge~\cite{suppl}.
A new insight  for its ground state could be a  gapless spin liquid with SFSs but preserving TRS, which we leave for future work.

{\it Summary.} We have studied the spin-$1/2$ $J_1$-$J_2$-$J_3$ THM by extensive DMRG calculations.
We identify a CSL state spontaneously breaking TRS, featuring a chiral edge mode and spin pumping upon flux insertion.
The vanishing spin triplet gap and finite central charge reveal the gapless nature of this state.
The central charge which grows with system circumference further indicates emergent SFSs.
While the competing $J_2, J_3$ couplings lead to a gapped CSL on kagome lattice~\cite{he2014csl, gong2015}, they induce a gapless CSL on triangular lattice.
On the mean-field level we propose a staggered flux state driven by the $J_2, J_3$ couplings, which breaks TRS and forms SFS, providing a theoretical understanding for such a novel gapless phase.
The discovery of this gapless CSL reveals the novel possibility for the coexistence of chiral edge modes and SFSs in a gapless QSL, emergent from competing interactions in a frustrated two-dimensional magnet. 

{\it Note added.} After completion of this work, we became aware of a work by Shijie Hu et
al.~\cite{hu2019}, who studied the $J_1$-$J_2$ spin liquid. Compared to their work, our work focuses on the $J_1$-$J_2$-$J_3$ model and found a staggered flux state driven by further-neighbor interactions.

\acknowledgements
We would like to thank Olexei I. Motrunich, Tao Li, and Yuan Wan for extensive discussions. S.S.G. is supported by the National Natural Science Foundation of China Grants (11874078, 11834014) and the Fundamental Research Funds for the Central Universities. M.L. is supported by the National Science Foundation through PREM Grant DMR-1828019. D.N.S. is supported by the U.S. Department of Energy, Office of Basic Energy Sciences under Grant No. DE-FG02-06ER46305. W.Z. and Y.M.L. are supported by National Science Foundation under award number DMR-1653769.

\setcounter{figure}{0}
\makeatletter
\onecolumngrid

\begin{center}
\begin{Large}
{\bf Supplemental Material}
\end{Large}
\end{center}
\section{Ground-state energy}
\label{sup:energy}

\begin{figure}[t]
\includegraphics[width = 0.5\linewidth]{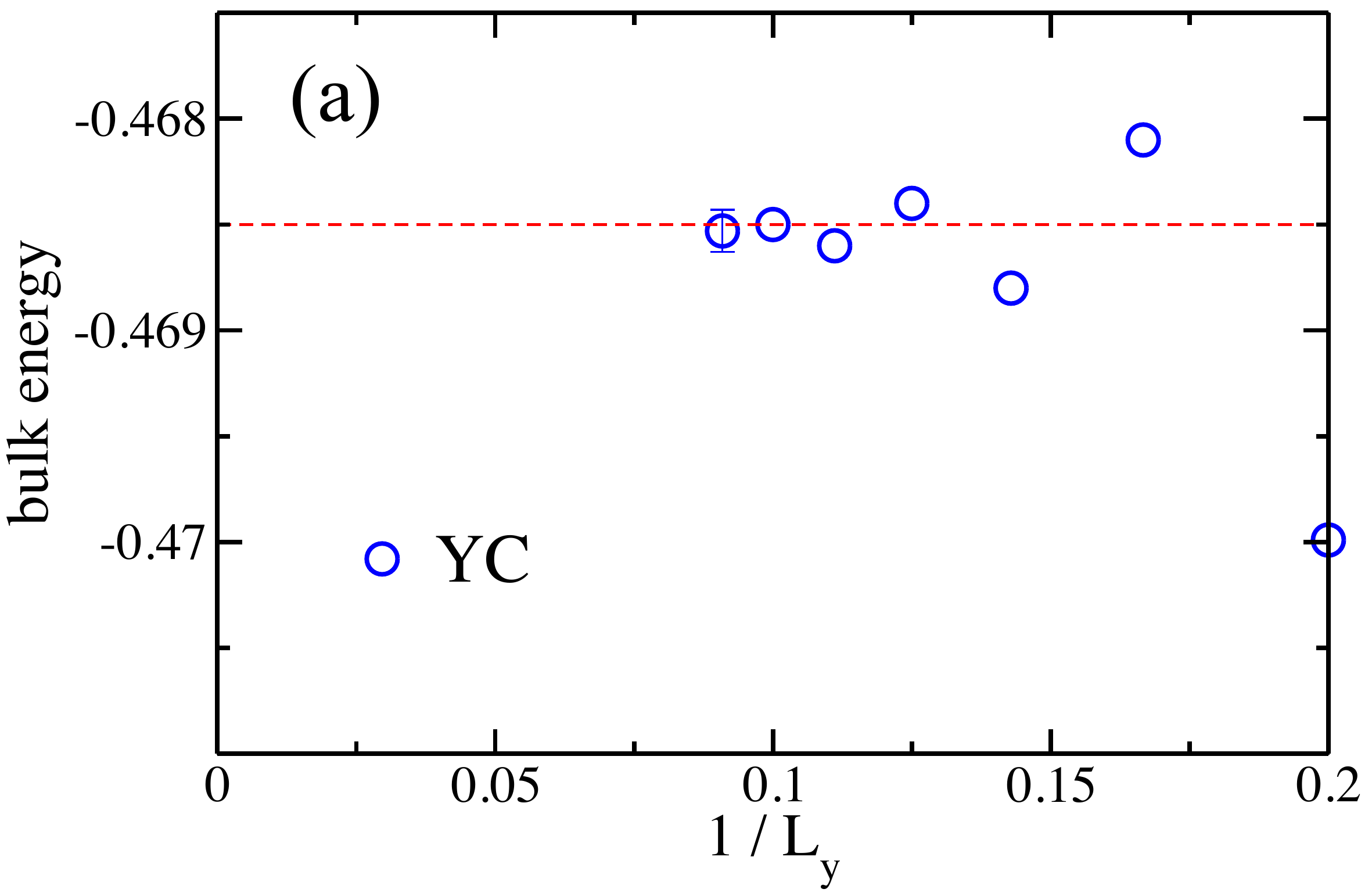}
\includegraphics[width = 0.48\linewidth]{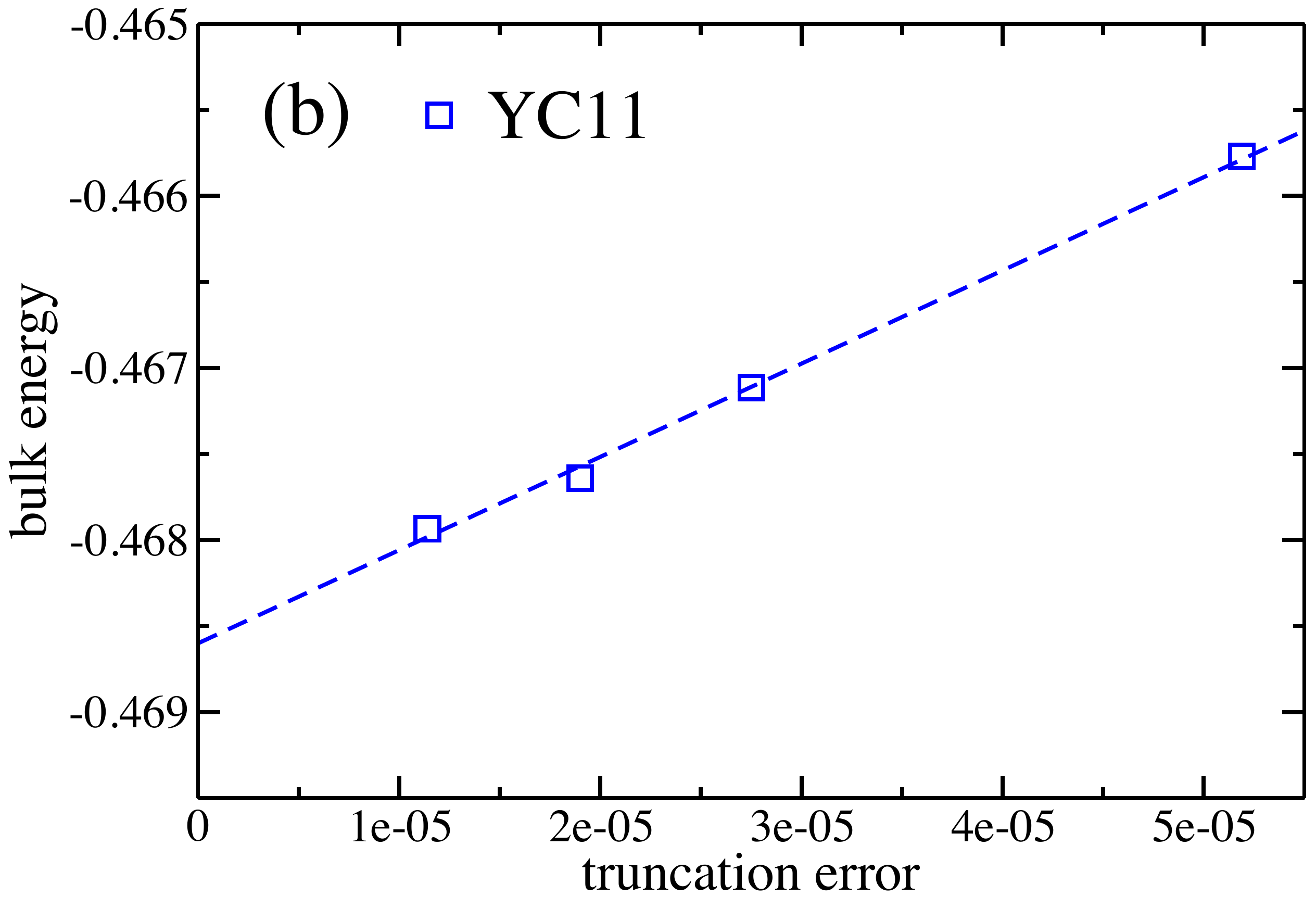}
\caption{Ground-state energy per site in the bulk of the YC cylinder for $J_2 = 0.3, J_3 = 0.15$. (a) The bulk energy for $L_y = 5 - 11$. The error bar for the energy of the YC11 cylinder is from the estimation of the energy in the extrapolation versus DMRG truncation error as shown in (b). The dashed red line in (a) denotes the energy $-0.4685$ as a rough estimation of the ground-state energy in the thermodynamic limit.
}
\label{figsup:energy}
\end{figure}

We first demonstrate the bulk energy on the YC cylinder in the gapless chiral spin liquid phase by choosing the parameter point at $J_2 = 0.3, J_3 = 0.15$ as a representative.
We show the energy versus cylinder circumference for $L_y = 5 - 11$ in Fig.~\ref{figsup:energy}(a).
On the YC5 cylinder, the ground state is a dimer state that breaks lattice translational symmetry.
On the other YC cylinders, the bond energy is translational invariant in the bulk of cylinder.
One may find that the energy shows an even-odd oscillation versus $L_y$ but changes small, approaching to the same value with increasing circumference.
By a rough estimation, we propose the ground-state energy in the thermodynamic limit near $-0.4685$ for this parameter point.

\section{Zigzag order and incommensurate order}
\label{sup:order}

\begin{figure}[t]
\includegraphics[width = 0.48\linewidth]{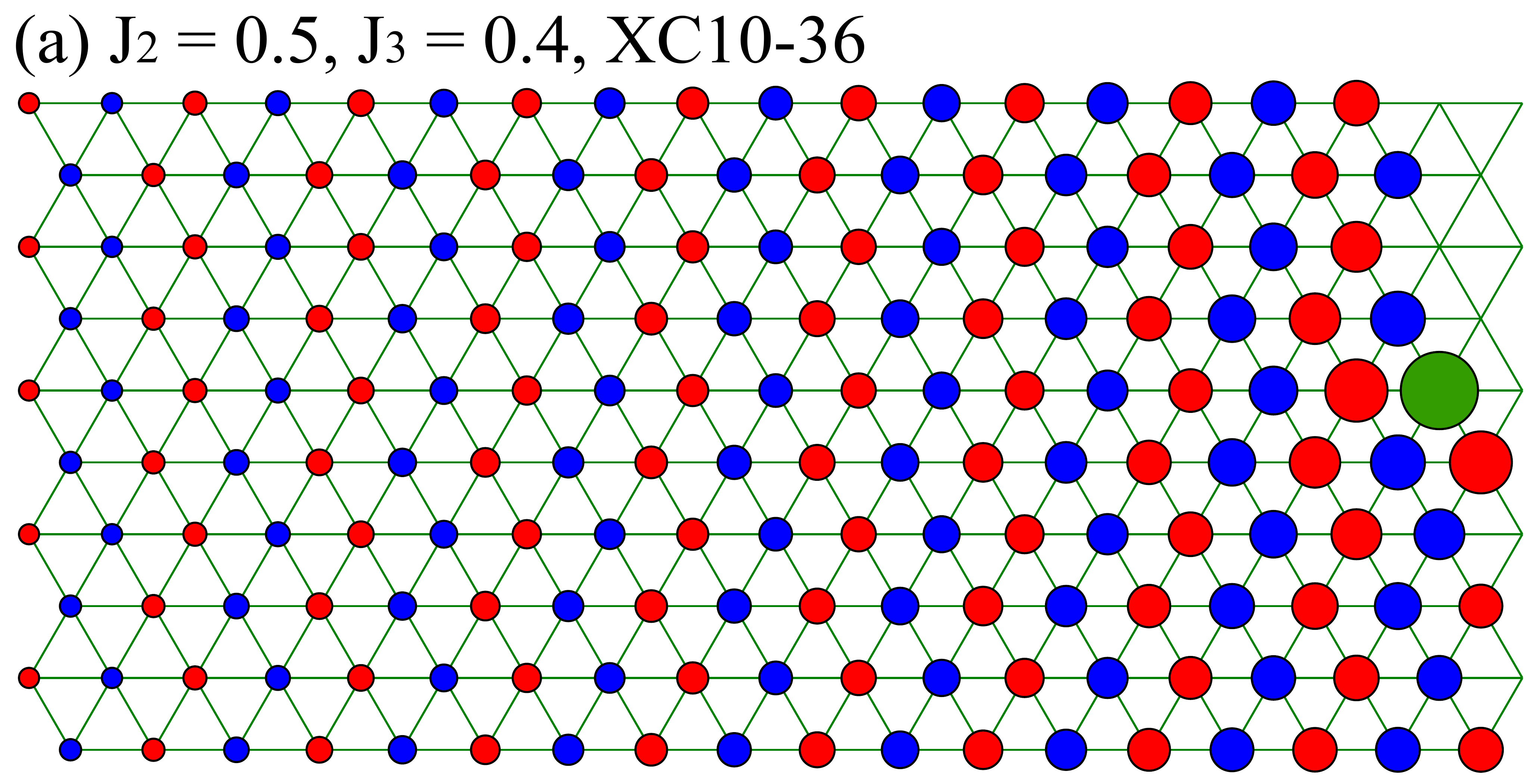}
\includegraphics[width = 0.48\linewidth]{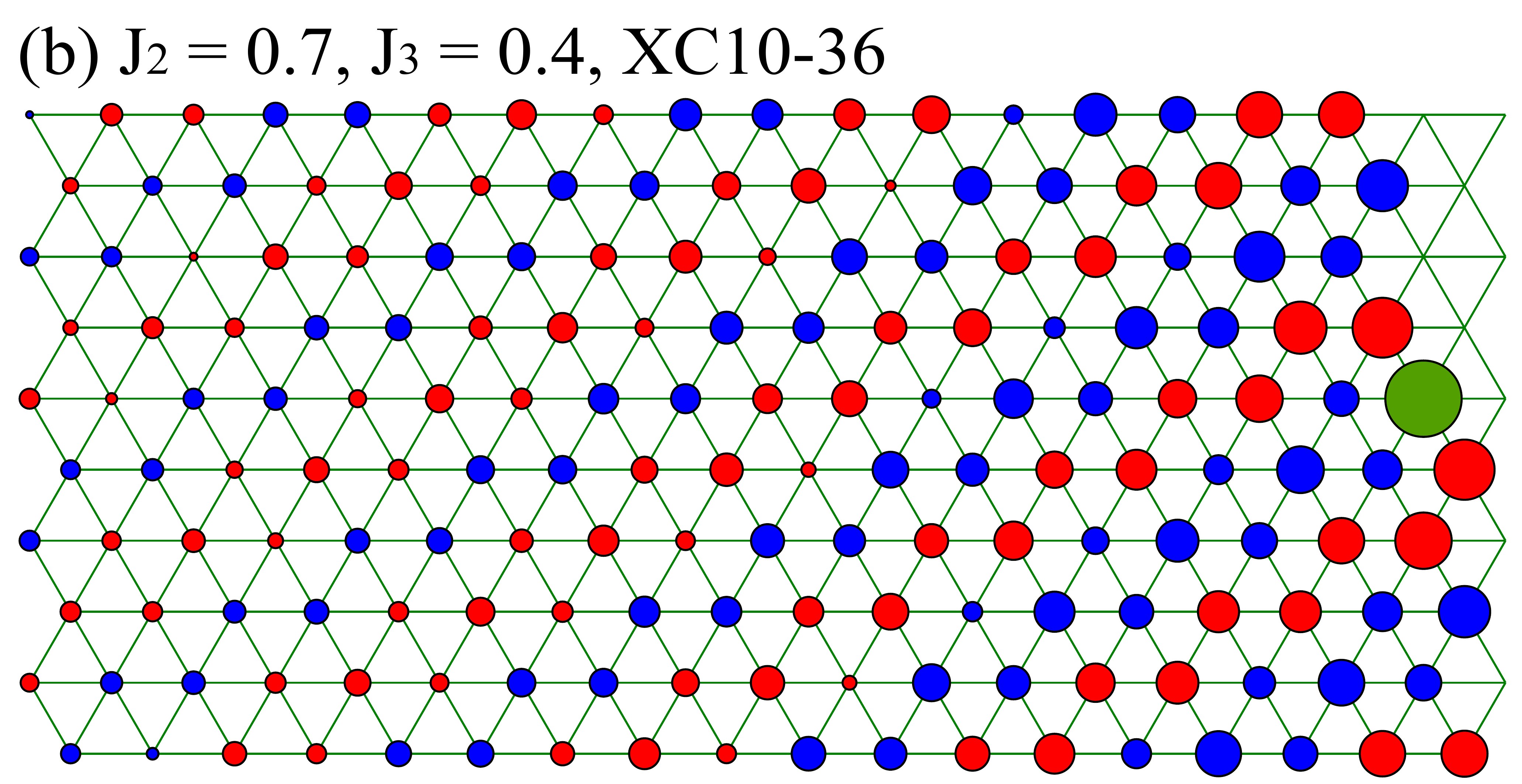}
\caption{Spin correlation functions on the XC10-36 cylinders for (a) $J_2 = 0.5, J_3 = 0.4$ in the zigzag order and (b) $J_2 = 0.7, J_3 = 0.4$ in the incommensurate order.
Here we just show the left half lattice.
The green site is the reference site in the middle of cylinder.
The blue and red circles denote the positive and negative spin correlations with respect to the reference site.
}
\label{figsup:order}
\end{figure}

We have shown the quantum phase diagram of the model in the main text, which has different magnetic order phases.
Besides the $120^0$ order and the stripe order which have been found in the $J_1 - J_2$ triangular Heisenberg model, we find another two new orders.
One order is called as zigzag order in our paper, as shown in Fig.~\ref{figsup:order}(a).
We have confirmed this order on both YC and XC cylinders. Here we just show the results on the XC cylinder.
The spins along the zigzag path in each column are parallel to each other.

Another order seems to be an incommensurate magnetic order.
As shown in Fig.~\ref{figsup:order}(b), the magnitudes of the magnetic moments have oscillation between large and small values.
The overall pattern of spin correlations seems compatible with the zigzag order but with oscillation, which suggests this order as the zigzag order with incommensurate modulation.
On the small XC cylinders such as the XC6 and XC8 cylinders, the spin correlations decay fast, and the order becomes stable on the larger XC10 cylinder.
On the YC cylinder, this incommensurate order is also found.

\section{Absent spin and dimer order in the gapless chiral spin liquid state}
\label{sup:spin_dimer}

\begin{figure}[t]
\includegraphics[width = 0.31\linewidth]{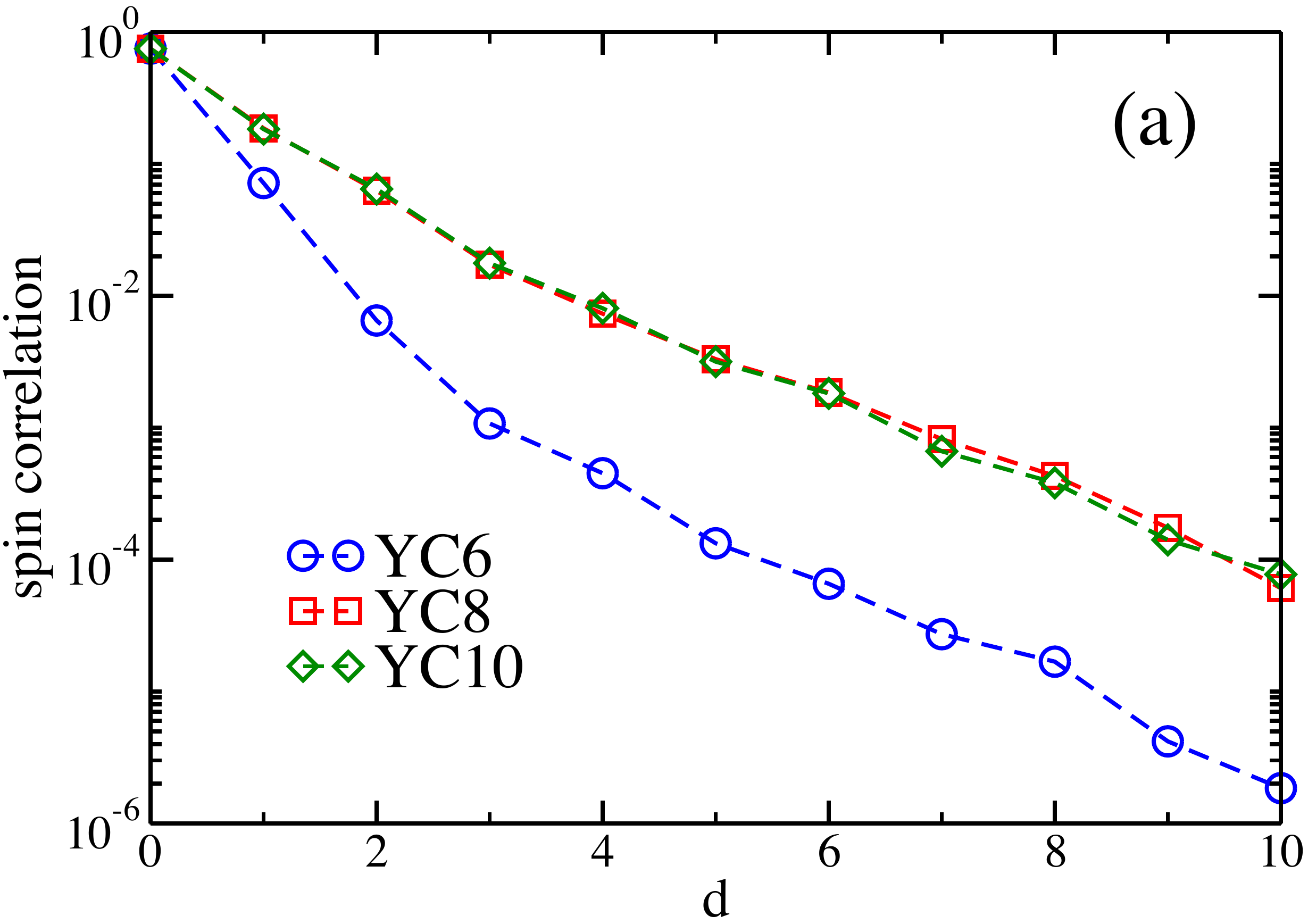}
\includegraphics[width = 0.18\linewidth]{sq_csl.pdf}
\includegraphics[width = 0.18\linewidth]{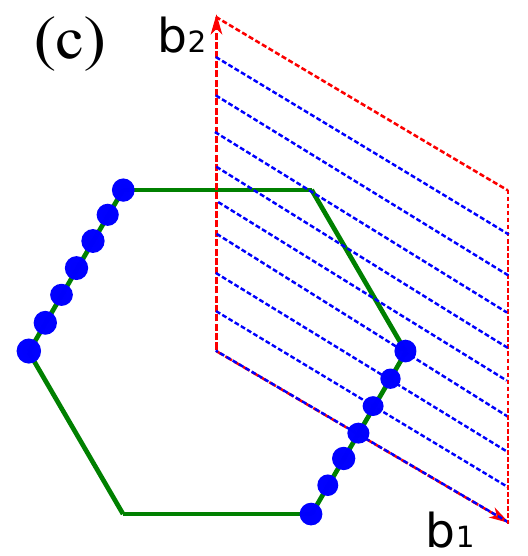}
\includegraphics[width = 0.31\linewidth]{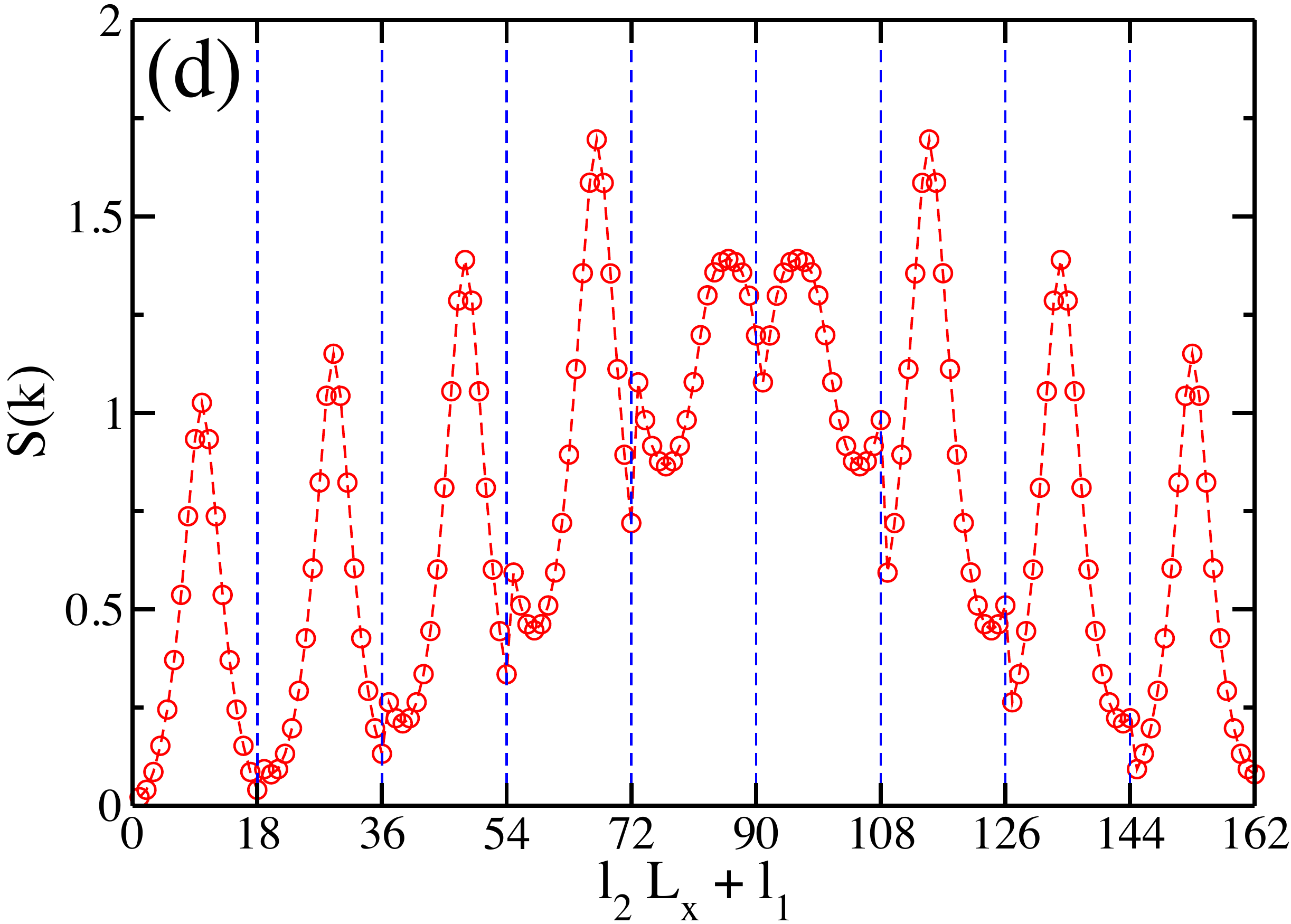}
\caption{Spin correlation function and spin structure factor for $J_2 = 0.3, J_3 = 0.15$ on the YC cylinder. (a) Log-linear plot of spin correlation function along the extended direction of cylinder. The reference site is in the middle of cylinder. (b) Spin structure factor on the YC9 cylinder. The dashed line denotes the Brillouin zone. (c) The Brillouin zone of the triangular lattice with the reciprocal vectors ${\bf b}_1$ and ${\bf b}_2$. The dashed lines denote the momentum points of the YC9 cylinder in the first Brillouin zone. The blue dots mark the small peaks of $S({\bf k})$ shown in (d). (d) Spin structure factor $S({\bf k})$ along the momentum lines in the first Brillouin zone in (c). The momentum positions of the peaks are denoted as blue dots in (c).
}
\label{figsup:spin_decay}
\end{figure}

In Fig.~\ref{figsup:spin_decay}, we show spin correlation function and structure factor of the gapless chiral spin liquid state.
We show the results for $J_2 = 0.3, J_3 = 0.15$.
As shown in Fig.~\ref{figsup:spin_decay}(a), one can find that spin correlation functions decay with distance like an exponential way, showing the absent magnetic order.
We define spin structure factor as
\begin{equation}
S({\bf k}) = \frac{1}{N} \sum_{i, j} \langle {\bf S}_i \cdot {\bf S}_j \rangle e^{i {\bf k} \cdot ({\bf r}_i - {\bf r}_j)}.
\end{equation}
In Fig.~\ref{figsup:spin_decay}(b), we show spin structure factor $S({\bf k})$ on the YC9-18 cylinder, which shows broad peaks near the corners of the Brillouin zone.
To explicitly show spin structure factor, we plot $S({\bf k})$ along the momentum lines in the first Brillouin zone in Fig.~\ref{figsup:spin_decay}(d).
The momentum lines are denoted by the dashed lines in Fig.~\ref{figsup:spin_decay}(c).
The $x$ label of Fig.~\ref{figsup:spin_decay}(d) $l_2 L_x + l_1$ denotes that the momentum points follow the sequence from left to right and from bottom to top in the first Brillouin zone in Fig.~\ref{figsup:spin_decay}(c).
The momentum positions of the small peaks are marked as blue dots in Fig.~\ref{figsup:spin_decay}(c).
One can find that there is no sharp peak of $S({\bf k})$, consistent with absent magnetic order. While the dominant peaks appear at the corners of the Brillouin zone, some smaller peaks appear along one edge of the Brillouin zone.
On the other YC cylinders such as YC8 and YC10, spin structure factors are similar to that of the YC9 cylinder.
The dominant peaks also appear at the corners of the Brillouin zone, but the smaller peaks may not exactly locate along the edge of Brillouin zone but have small deviation duo to the geometric size.

\begin{figure}[t]
\includegraphics[width = 0.2\linewidth]{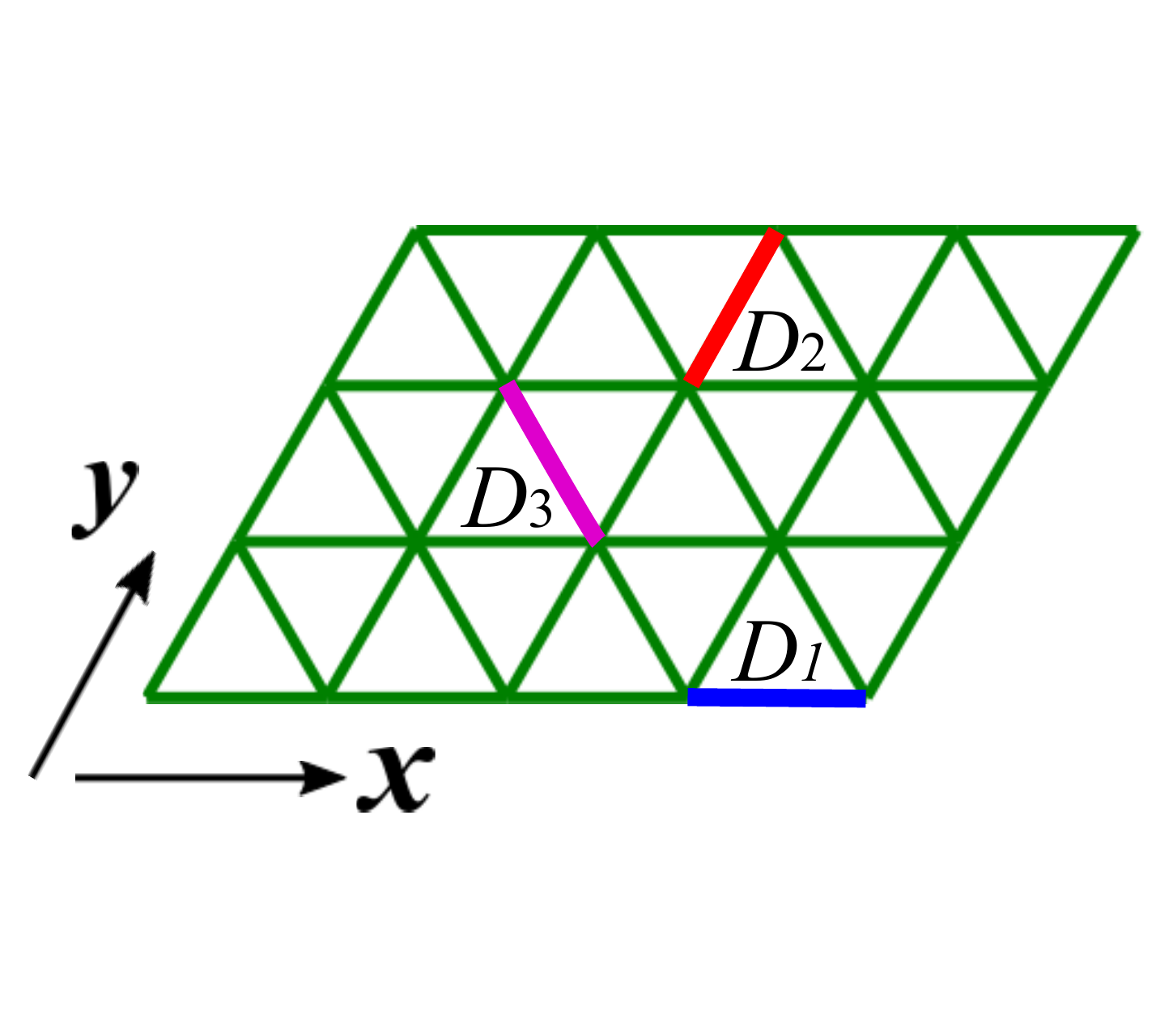}
\includegraphics[width = 0.26\linewidth]{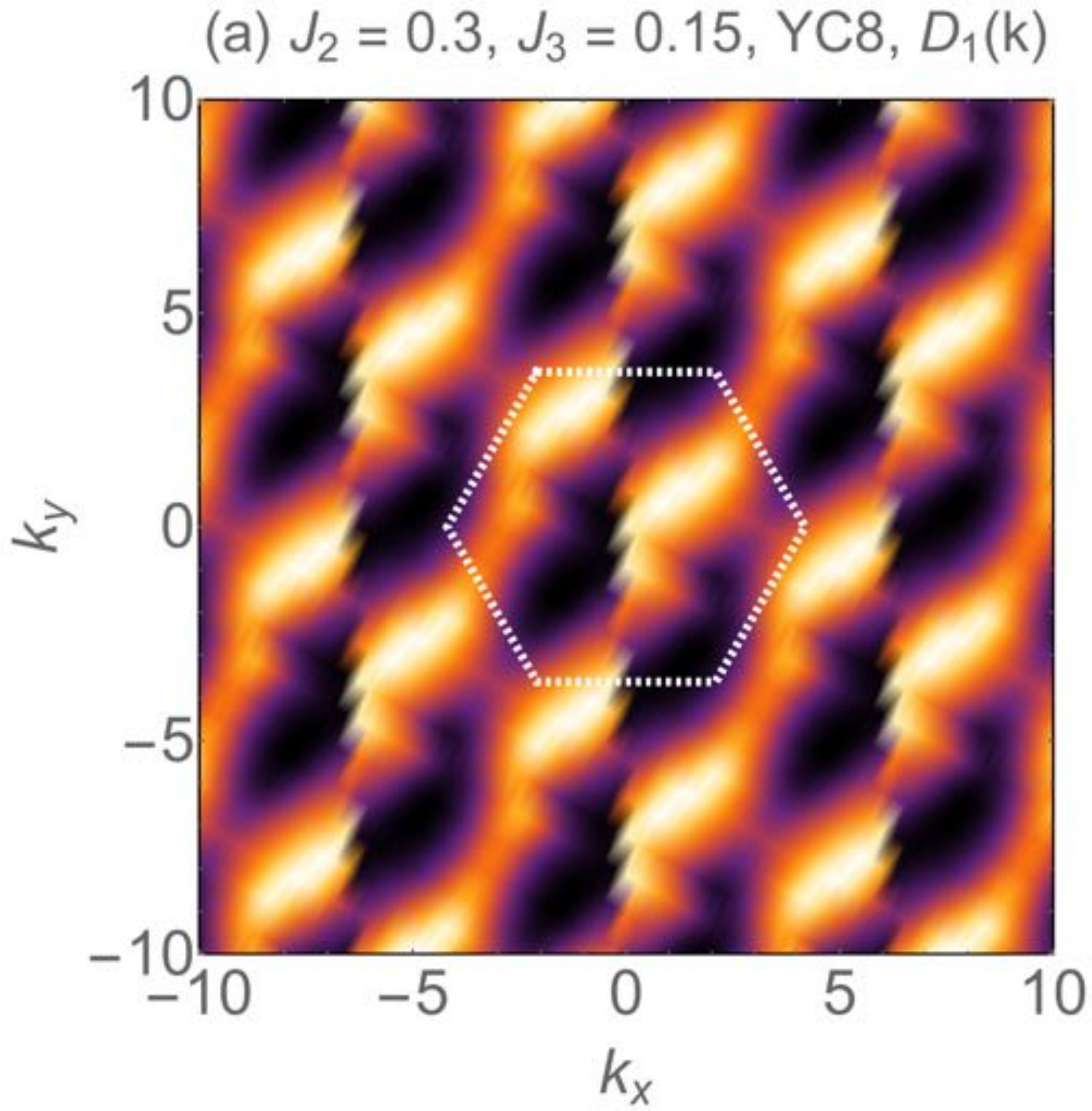}
\includegraphics[width = 0.26\linewidth]{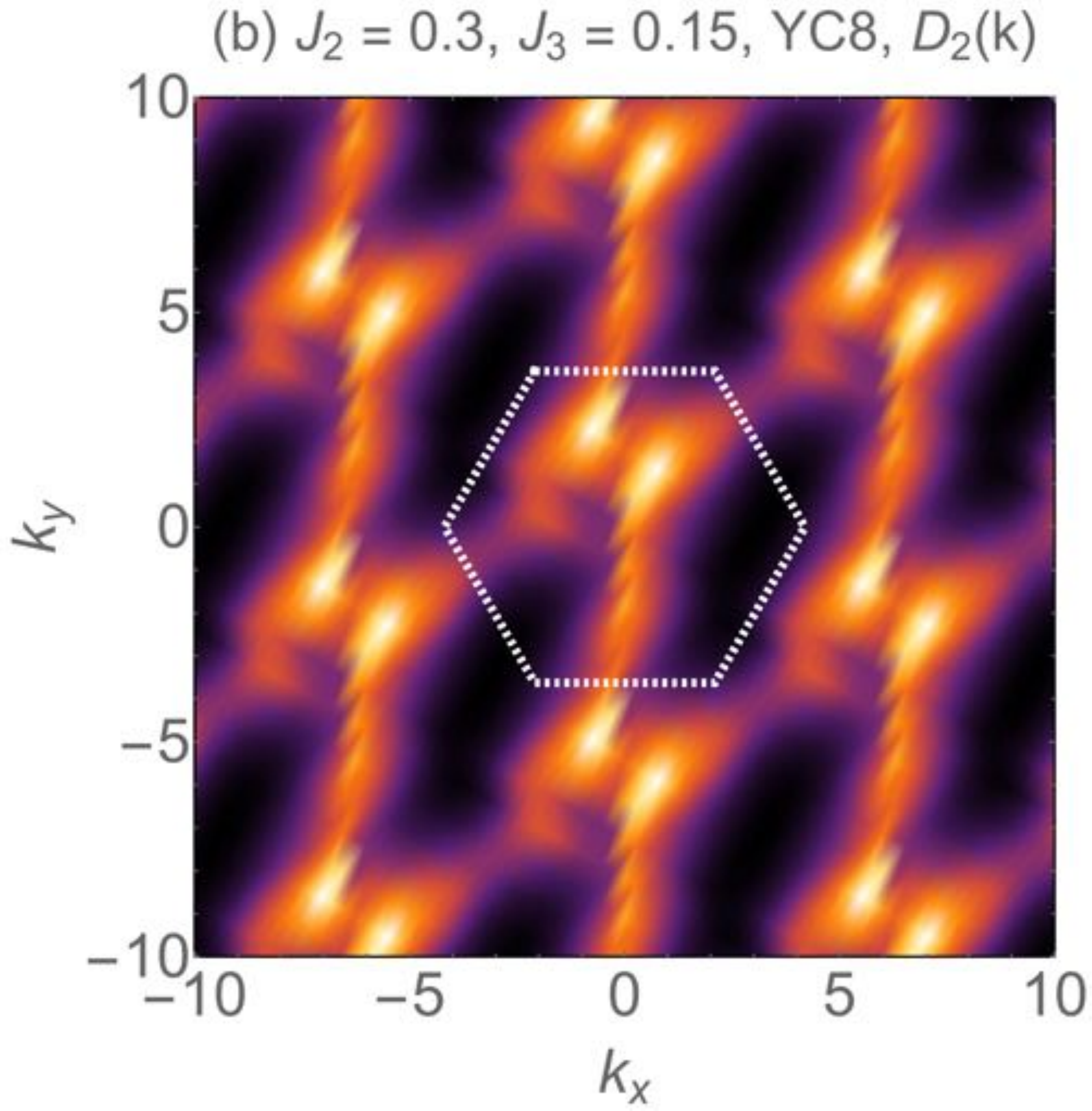}
\includegraphics[width = 0.26\linewidth]{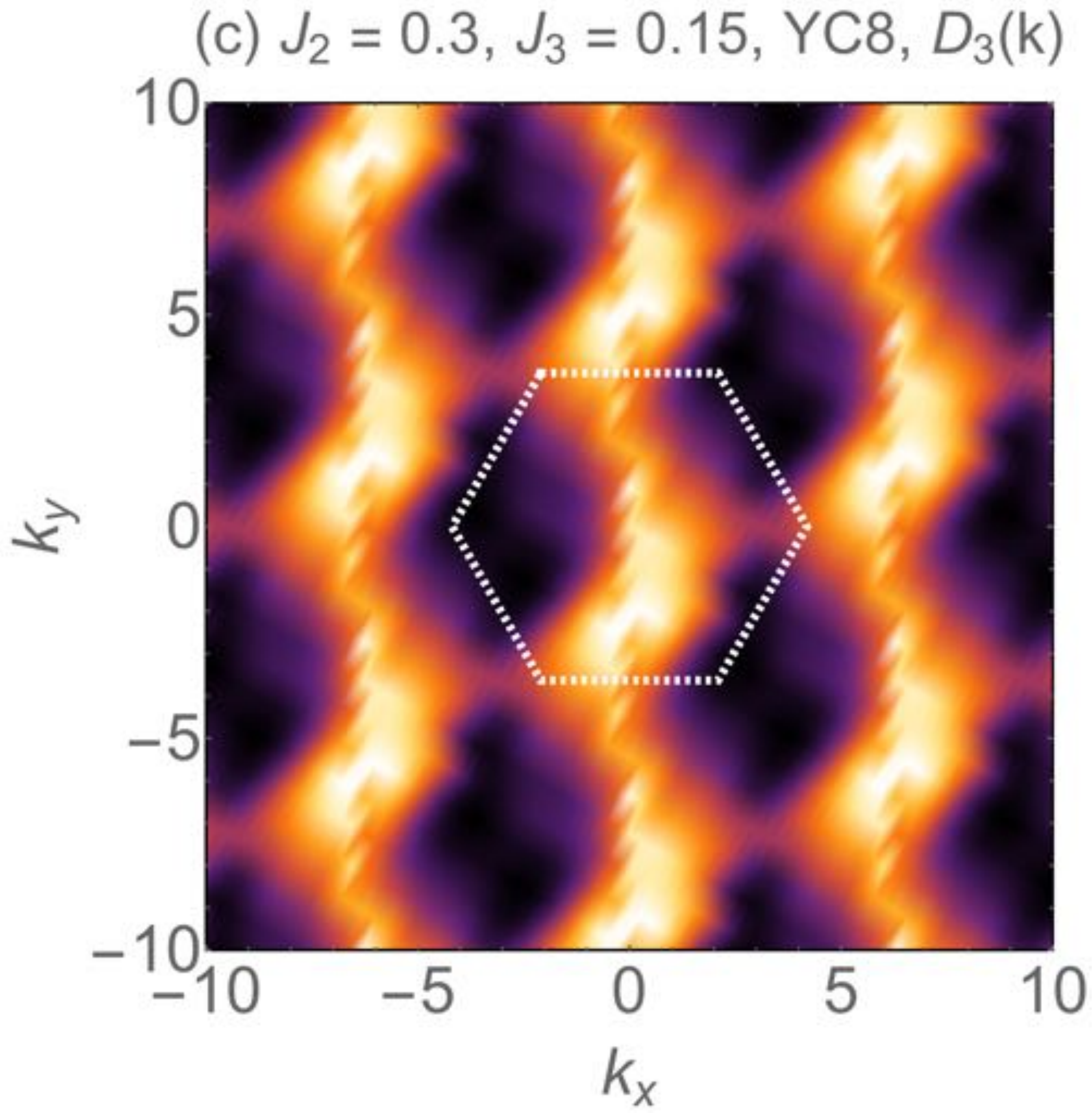}
\caption{Dimer structure factor on the YC8-24 cylinder for $J_2 = 0.3, J_3 = 0.15$.
The left insert shows the three directions of the dimer-dimer correlations defined in our calculation.
(a) - (c) are the dimer structure factors of the parallel dimer bonds along the three directions defined in the insert.
The white hexagon denotes the Brillouin zone.
None of the dimer structure factors has a sharp peak.}
\label{figsup:dimer_sq}
\end{figure}

To study dimer order in the spin liquid phase, we calculate dimer correlation functions and their structure factors.
Since the nearest-neighbor bonds on the triangular lattice are along three different directions as shown in the inset of Fig.~\ref{figsup:dimer_sq}, we calculate their dimer structure factors separately.
The dimer structure factors $D_{l}({\bf k})$ $(l = 1, 2, 3)$ can be defined as
\begin{equation}
D_{l}({\bf k}) = \frac{1}{N} \sum_{i,j} (\langle \mathcal{B}_{i, i+\delta_l} \mathcal{B}_{j, j+\delta_l} \rangle - \langle \mathcal{B}_{i, i+\delta_l} \rangle \langle \mathcal{B}_{j, j+\delta_l} \rangle) e^{i {\bf k} \cdot ({\bf r}_i - {\bf r}_j)},
\end{equation}
where $\mathcal{B}_{i, i+\delta_l} = {\bf S}_i \cdot {\bf S}_{i+\delta_l}$ with $\delta_l$ denoting the three bond directions $\delta_1 = (1, 0)$, $\delta_2 = (\frac{1}{2}, \frac{\sqrt{3}}{2})$, $\delta_3 = (-\frac{1}{2}, \frac{\sqrt{3}}{2})$.
The obtained dimer structure factors on different YC cylinders such as YC8, YC9, and YC10 are similar.
Here we show the results on the YC8 cylinder in Fig.~\ref{figsup:dimer_sq} as an example.
The dimer structure factors along the three directions are all featureless, showing the vanished dimer order in the spin liquid phase.

\subsection{Size scaling of spin triplet gap}

\begin{figure}[t]
\includegraphics[width = 0.5\linewidth]{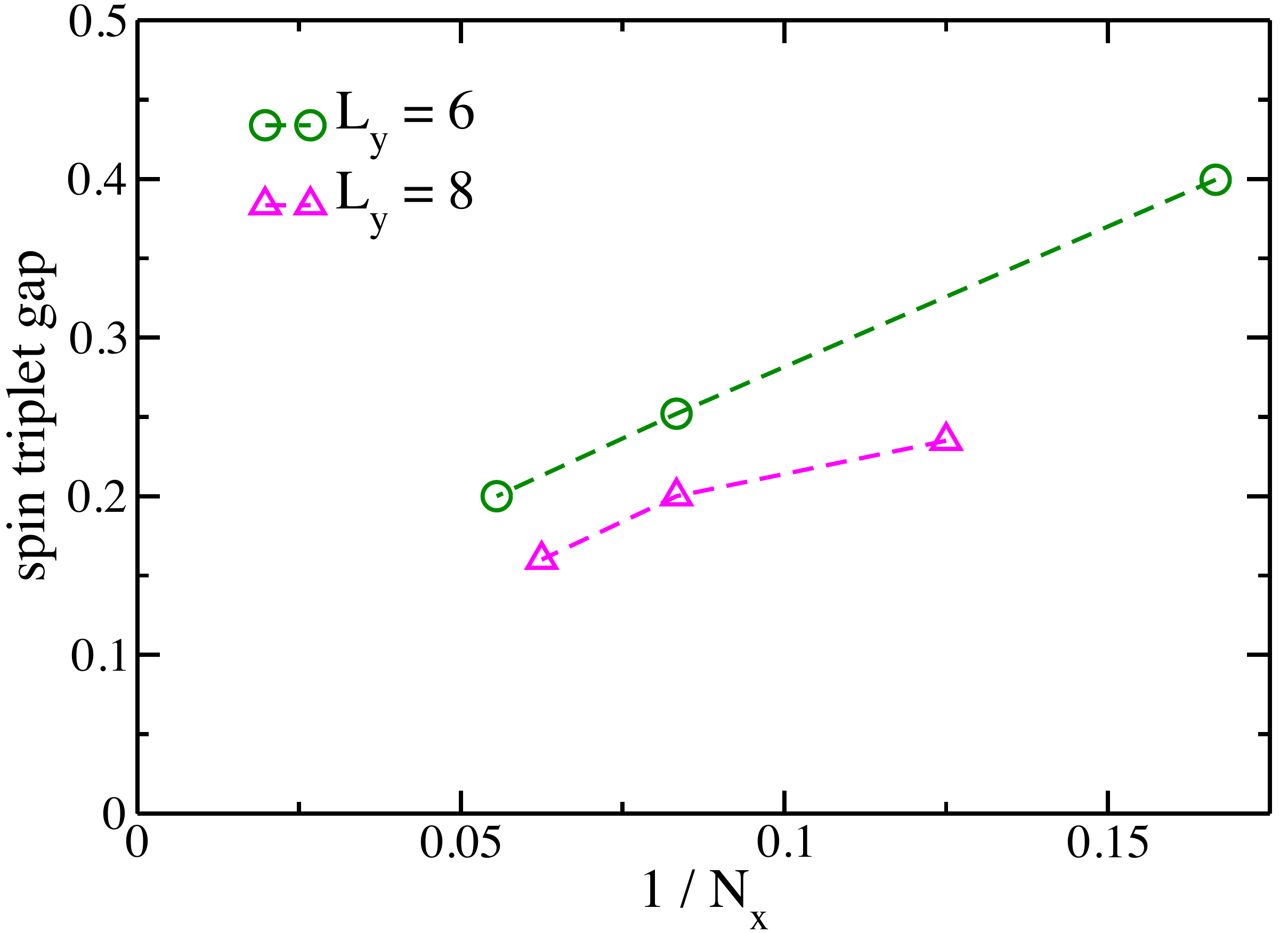}
\caption{Size scaling of spin triplet gap on the YC6 and YC8 cylinders for $J_2 = 0.3, J_3 = 0.15$.
We obtain the ground state on a long cylinder and calculate the gap for the middle $N_x \times L_y$ sites in the bulk of cylinder.}
\label{figsup:gap}
\end{figure}

In the main text, we have shown the $1 / L_y$ size scaling of the spin triplet gap obtained on the square-like clusters for $J_2 = 0.3, J_3 = 0.15$.
Here we demonstrate the $1 / N_x$ size scaling of the gap on the YC6 and YC8 cylinders.
We calculate the ground state on a long cylinder and then sweep the middle $N_x \times L_y$ sites in the bulk of cylinder by targeting the total spin $S = 1$ sector.
The obtained spin gap versus $1 / N_x$ is shown in Fig.~\ref{figsup:gap}.
We find that the gap depends on the system length and seems to decrease with $N_x$, which is consistent with the 1D gapless feature.
To avoid the 1D physics, we use the square-like clusters for $1 / L_y$ size scaling to estimate the spin triplet gap in the thermodynamic limit.

\subsection{Entanglement spectrum and chiral edge mode}

\begin{figure}[t]
\includegraphics[width = 0.6\linewidth]{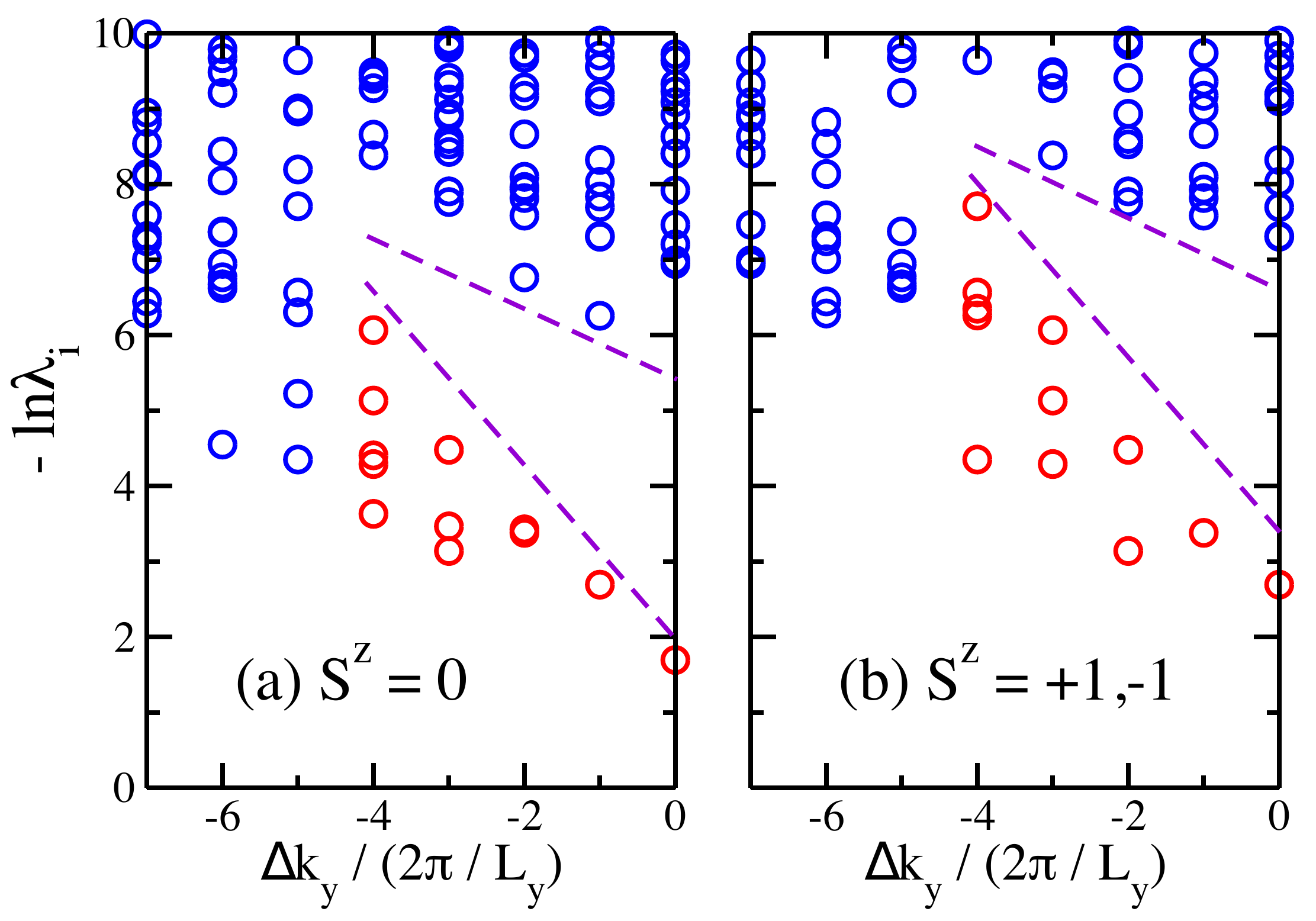}
\caption{Entanglement spectrum for $J_2 = 0.3, J_3 = 0.15$ on the YC8 cylinder.
The entanglement spectrum is labeled by the quantum numbers total spin $S^z = 0$ (a),  $\pm 1$ (b) and relative momentum along the $y$ direction $\Delta k_y$.
$\lambda_i$ is the eigenvalue of reduced density matrix.
The red circles denote the near degenerate pattern $\{1, 1, 2, 3, 5, \cdots \}$ of the low-lying spectrum with different quantum numbers $\Delta k_y$.
}
\label{figsup:spectrum}
\end{figure}

Next, we study the entanglement spectrum in the gapless chiral spin liquid state.
In chiral topological states, it was conjectured that there is a one-to-one correspondence between entanglement spectrum and physical edge spectrum~\cite{li2008}.
In Fig.~\ref{figsup:spectrum}, we show the entanglement spectrum on the YC8 cylinder, which is labeled by the momentum quantum number along the ${\bf a}_2$ direction $\Delta k_y$ in different total $S^z$ sectors.
The spectrum shows quasi-degenerate groups of levels with counting $\{ 1, 1, 2, 3, 5, \cdots \}$, which agrees with the chiral SU(2)$_1$ Wess-Zumino-Witten conformal field theory~\cite{cft}.

The spectrum result seems to suggest a gapless chiral spin liquid state with chiral edge mode.
Such chiral states with chiral edge mode and gapless bulk properties have been studied recently for non-interacting systems~\cite{dubail2015}, such as the non-interacting $p+ip$ topological chiral state.
However, for interacting systems, such example is rare.
This gapless chiral spin liquid state may be the first example to realize such a novel quantum state for interacting systems, which should be further studied in future work.

\subsection{Quantum phase transition from the $J_1 - J_2$ spin liquid phase to the gapless chiral spin liquid phase}
\begin{figure}[t]
\includegraphics[width = 0.48\linewidth]{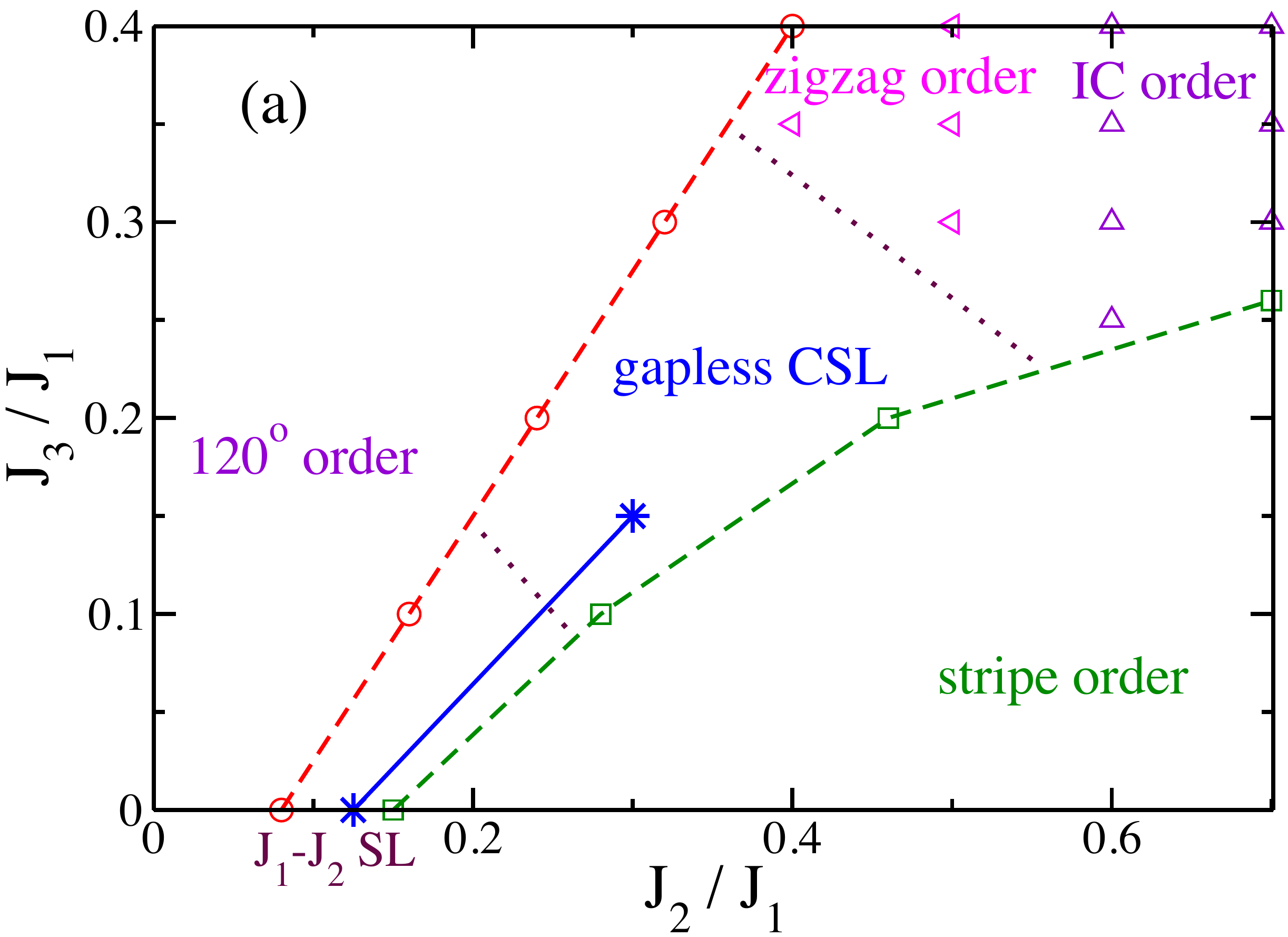}
\includegraphics[width = 0.48\linewidth]{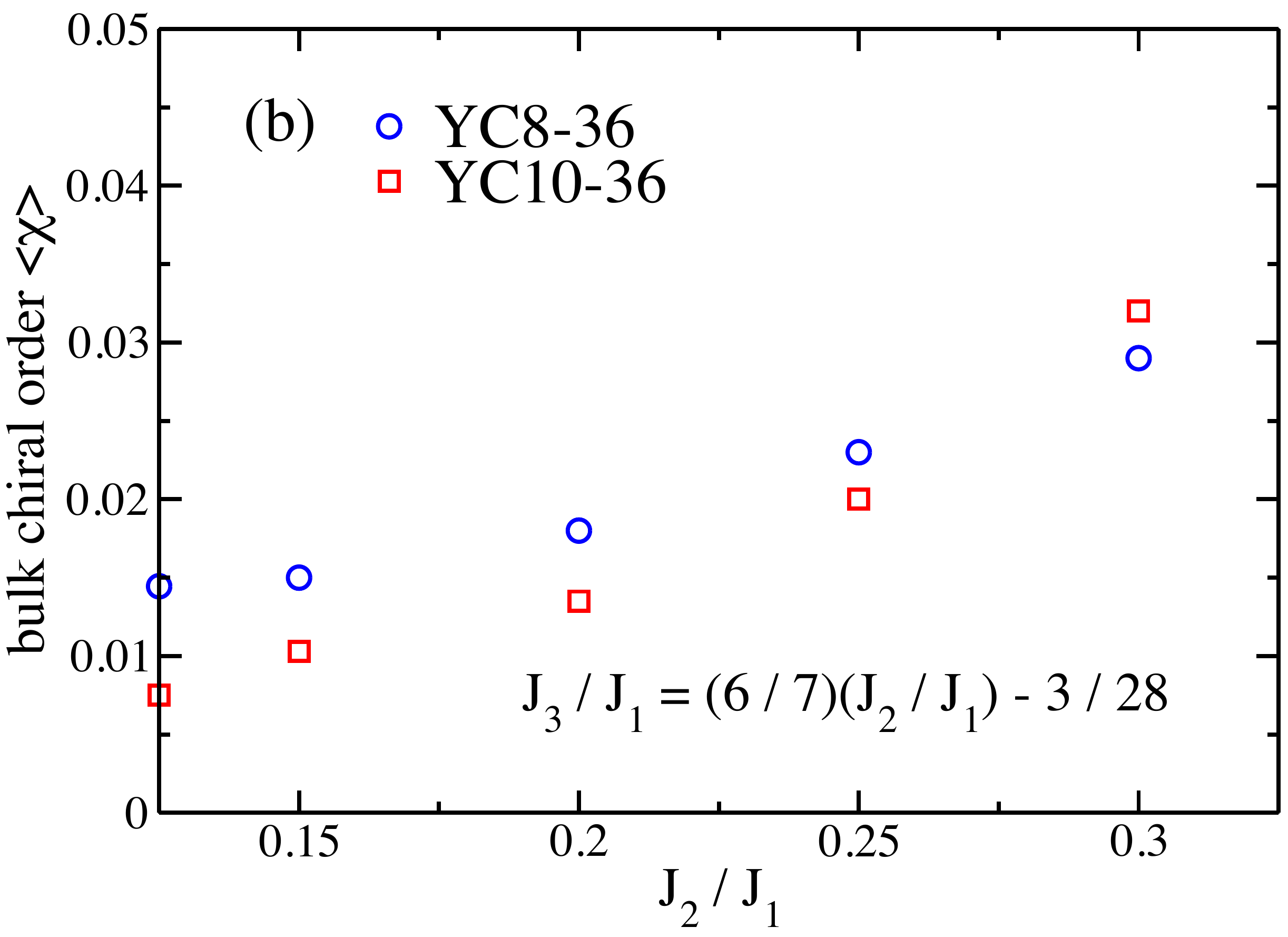}
\includegraphics[width = 0.48\linewidth]{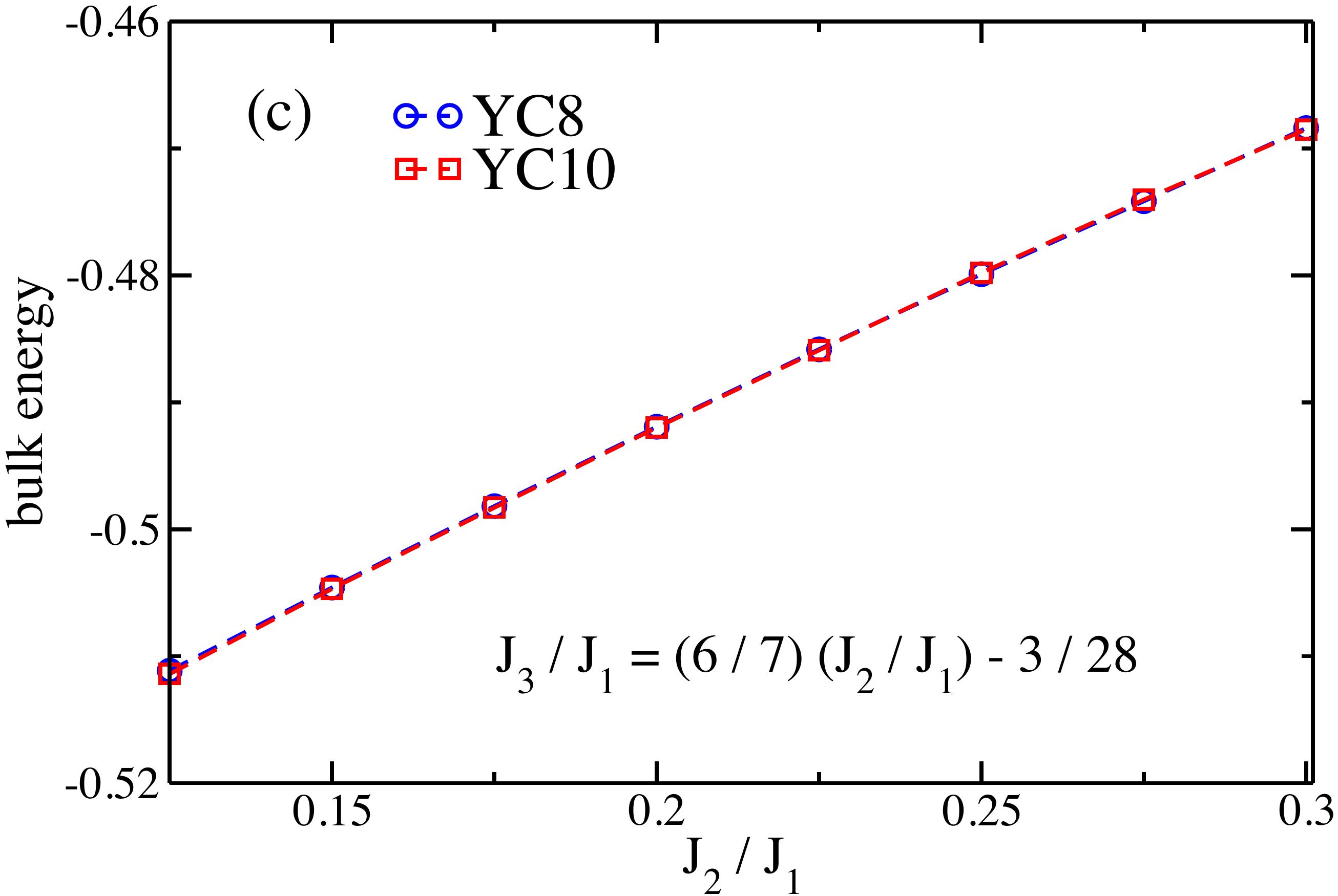}
\includegraphics[width = 0.46\linewidth]{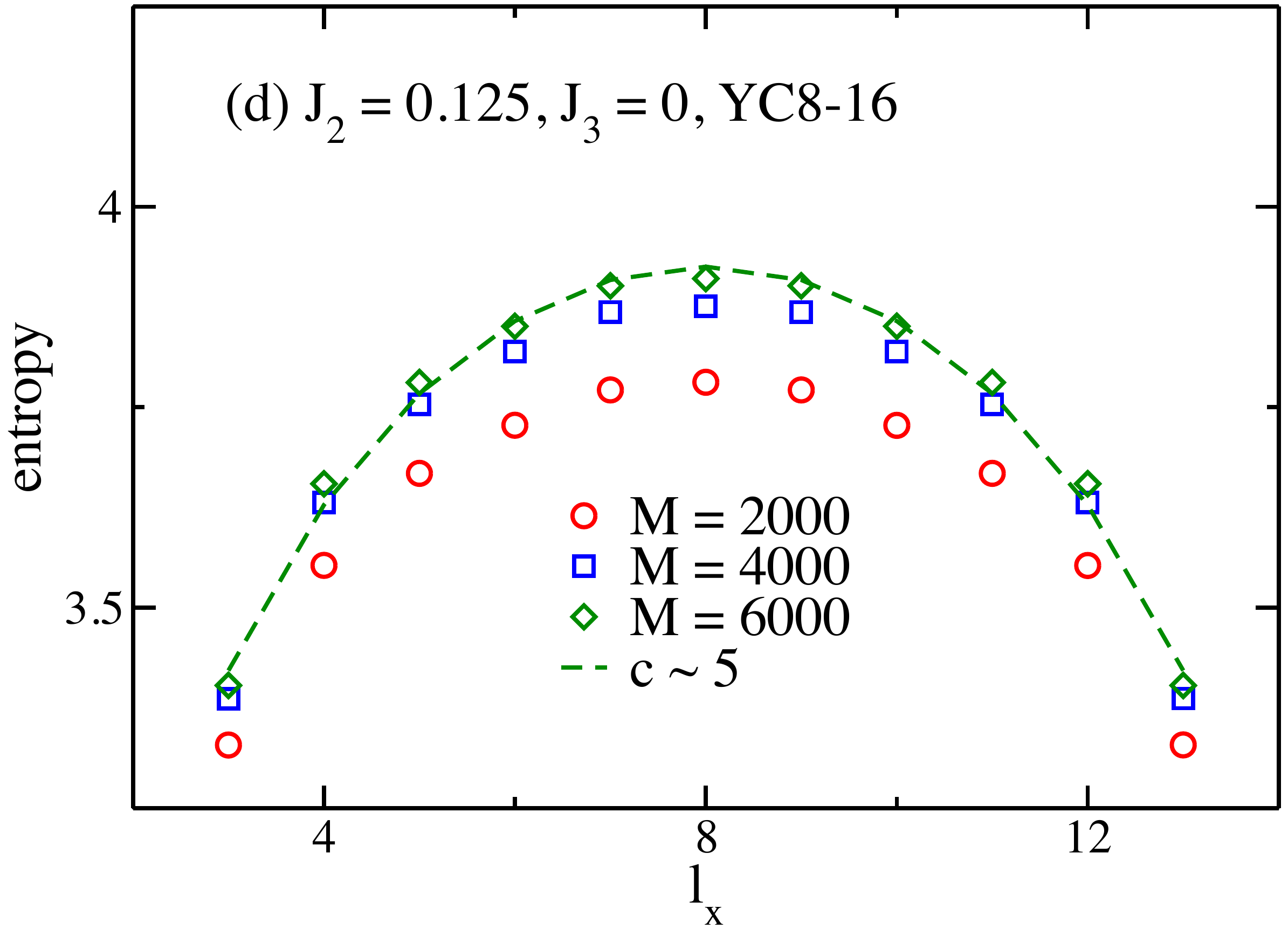}
\caption{Quantum phase transition from the $J_1 - J_2$ spin liquid phase to the gapless chiral spin liquid phase.
(a) Quantum phase diagram of the model with the blue solid line denoting the studied parameter line to show the phase transition. This line is defined as $J_3 / J_1 = (6/7) (J_2 / J_1) - 3 / 28$, which crosses the typical parameter points in the two phases $J_2 = 0.125, J_3 = 0$ and $J_2 = 0.3, J_3 = 0.15$.
Scalar chiral order (b) and ground-state energy (c) in the bulk of the YC8 and YC10 cylinders versus $J_2/J_1$ along the blue solid parameter line.
(d) Entanglement entropy versus subsystem length $l_x$ for $J_2 = 0.125, J_3 = 0$ on the YC8-16 cylinder by keeping different state numbers.
}
\label{figsup:transition}
\end{figure}

Since this gapless chiral spin liquid phase emerges at the neighbor of the $J_1 - J_2$ spin liquid phase, it would be very interesting to study how the system evolves from one state to the other one.
To study this phase transition, we choose a parameter line that crosses the typical parameter points in the two phases that are $J_2 = 0.125, J_3 = 0$ and $J_2 = 0.3, J_3 = 0.15$.
Thus the parameter line is defined as $J_3 / J_1 = (6/7) (J_2 / J_1) - 3 / 28$, which is shown in Fig.~\ref{figsup:transition}(a) as the blue solid line.
First of all, we calculate the scalar chiral order in the bulk of the YC8-36 and the YC10-36 cylinders along this parameter line as shown in Fig.~\ref{figsup:transition}(b).
While the chiral order decreases with circumference for $J_2 < 0.3$, it becomes stable with growing circumference at $J_2 = 0.3, J_3 = 0.15$, showing the absent and emergent scalar chiral order in either of the two phases.
Furthermore, we calculate the ground-state energies in the bulk of cylinder along the parameter line as shown in Fig.~\ref{figsup:transition}(c), which change very small from the YC8 to the YC10 cylinder.
With growing couplings, the energy seems very smooth, suggesting a possible continuous phase transition between the two phases although a weak first-order transition cannot be excluded.

Next we reexamine the entanglement entropy in the $J_1 - J_2$ spin liquid state.
A typical point $J_2 = 0.125, J_3 = 0$ is shown in Fig.~\ref{figsup:transition}(d) on the YC8-16 cylinder.
With growing number of kept states, the entropy keeps to increase.
We can see that the entropy also has a logarithmic correction of the area law.
The data by keeping $6000$ $SU(2)$ states could be fitted quite well by the formula $S(l_x) = (c/6) \ln[ (L_x / \pi) \sin(l_x \pi / L_x) ] + g$~\cite{calabrese2004}, where $S(l_x)$ is the bipartite entanglement entropy, $c$ is the central charge, and $g$ is a nonuniversal constant.
The fitting gives a finite central charge $c \sim 5$, which is consistent with the central charge of the point $J_2 = 0.3, J_3 = 0.15$ (in the gapless chiral spin liquid phase) on the same system size.
The similar behavior of entropy suggests possible spinon Fermi surface in the $J_1 - J_2$ spin liquid state.
A new insight for the ground-state of the $J_1 - J_2$ spin liquid could be a gapless spin liquid with small spinon Fermi surface, which however preserves time-reversal symmetry.
In the presence of the third-neighbor $J_3$ coupling, time-reversal symmetry is broken and the system evolves to the gapless chiral spin liquid phase.
This new insight needs more studies in future work.

\section{Details in the mean-field analysis for the staggered flux state}
The ground state is at the half-filling of fermionic spinons.
We know that the U(1) Dirac state does not break translational symmetry after projection.
However, on mean-field level, the unit cell has to be doubled due to magnetic translation~\cite{PhysRevLett.98.117205}.
Now therefore, we choose to label the lattice as ${\bf a}_1^{\prime} = (1, 0), {\bf a}_{2}^{\prime} = (1, \sqrt{3}), {\bf r} = r_{1} {\bf a}_{1}^{\prime} + r_{2} {\bf a}_{2}^{\prime}, r_{1}, r_{2} \in {\bf Z}$.
The dual reciprocal space are labeled as $\mathbf{b}_{1}^{\prime} = (1, -\sqrt{3}/3)$, $\mathbf{b}_{2}^{\prime} = (0, \sqrt{3}/3), \mathbf{k}= k_{1}\mathbf{b}_{1}^{\prime} + k_{2}\mathbf{b}_{2}^{\prime}$, $k_{1} = 2\pi l_{1} / N_{1}$, $k_{2} = 2\pi l_{2} / N_{2}$, $l_1, l_2 \in {\bf Z}$, where $N_{1}, N_{2}$ are the numbers of the unit cells along ${\bf a}_{1}^{\prime}$- and ${\bf a}_{2}^{\prime}$-direction, respectively.
The first Brillouin zone (FBZ) can be identified as $k_{1}, k_{2}\in[-\pi, \pi)$.
We also define the three NN bridge vectors in Descartes coordinate system:
\begin{equation}
\begin{aligned}
    \delta_{1}&=(1, 0) = {\bf a}_1^{\prime}, \\
    \delta_{2}&=\left(\frac{1}{2}, \frac{\sqrt{3}}{2}\right) = \frac{{\bf a}_2^{\prime}}{2}, \\
    \delta_{3}&=\left(-\frac{1}{2}, \frac{\sqrt{3}}{2}\right) = \frac{{\bf a}_2^{\prime}}{2} - {\bf a}_1^{\prime}.
\end{aligned}
\end{equation}
In the next we only consider one of the two branches of U(1) spin liquid fermionic spinons, which are spinless.
First of all, we consider the mean-field Hamiltonian with the NN hopping terms according to Eq. (5) of the main text, which is given as
\begin{equation}
\begin{aligned}
     H_{1}
     &=\chi\sum_{\mathbf{r}}
     \left(e^{\text{i}\phi_{1}}f_{\mathbf{r}}^{\dagger}f_{\mathbf{r}+\delta_{1}}+e^{-\text{i}\phi_{1}}f_{\mathbf{r}+\delta_{2}}^{\dagger}f_{\mathbf{r}+\delta_{1}+\delta_{2}} \right. \\
     &+e^{\text{i}\phi_{2}}f_{\mathbf{r}}^{\dagger}f_{\mathbf{r}+\delta_{2}}-e^{-\text{i}\phi_{2}}f_{\mathbf{r}+\delta_{2}}^{\dagger}f_{\mathbf{r}+2\delta_{2}} \\
     &\left.+f_{\mathbf{r}}^{\dagger}f_{\mathbf{r}+\delta_{3}}+f_{\mathbf{r}+\delta_{2}}^{\dagger}f_{\mathbf{r}+\delta_{3}+\delta_{2}}+h.c.\right),
\end{aligned}
\label{eq:ham_h1}
\end{equation}
We set the NN hopping amplitude $\chi = 1.0$ as unit.
In reciprocal lattice with the Fourier transformation on the torus
\begin{equation}
    f_{\mathbf{r} p}=\frac{1}{\sqrt{N}}\sum_{\mathbf{k}}e^{\text{i}\mathbf{k}\cdot\mathbf{r}}f_{\mathbf{k} p},
    \end{equation}
we have
\begin{equation}
\begin{aligned}
    H_{1}
    &=2\chi\sum_{\mathbf{k}}\eta_{\mathbf{k}}^{\dagger}h^{(1)}(\mathbf{k})\eta_{\mathbf{k}}, \\
    h^{(1)}(\mathbf{k})
    &=\begin{bmatrix}
    h_{00}^{(1)}(\mathbf{k}) & h_{01}^{(1)}(\mathbf{k}) \\
    h_{10}^{(1)}(\mathbf{k}) & h_{11}^{(1)}(\mathbf{k})
    \end{bmatrix},
\end{aligned}
\end{equation}
where $\eta_{\mathbf{k}}=(f_{\mathbf{k}+}, f_{\mathbf{k}-})^{T}$ and
\begin{equation}
    \begin{aligned}
    h_{00}^{(1)}(\mathbf{k})&=\cos(\mathbf{k}\cdot\delta_{1}+\phi_{1}), \\
    h_{11}^{(1)}(\mathbf{k})&=\cos(\mathbf{k}\cdot\delta_{1}-\phi_{1}), \\
    h_{01}^{(1)}(\mathbf{k})&=\text{i}e^{\text{i}\phi_{2}}\sin(\mathbf{k}\cdot\delta_{2})+\cos(\mathbf{k}\cdot\delta_{3}), \\
    h_{10}^{(1)}(\mathbf{k})&=h_{01}^{(1)*}(\mathbf{k}).
    \end{aligned}
\end{equation}
We may assume real and equal amplitude $\lambda$ in the 2nd NN hopping, thus
\begin{equation}
\begin{aligned}
H_{2}
&=\lambda\sum_{\mathbf{r}}\left( e^{\text{i}\varphi_{1}}f_{\mathbf{r}}^{\dagger}f_{\mathbf{r}+\delta_{1}+\delta_{2}}+e^{-\text{i}\varphi_{1} }f_{\mathbf{r}+\delta_{2}}^{\dagger}f_{\mathbf{r}+\delta_{1}+2\delta_{2}} \right. \\
&+e^{\text{i}\varphi_{2}}f_{\mathbf{r}}^{\dagger}f_{\mathbf{r}+\delta_{2}+\delta_{3}}+e^{-\text{i}\varphi_{2} }f_{\mathbf{r}+\delta_{2}}^{\dagger}f_{\mathbf{r}+2\delta_{2}+\delta_{3}} \\
&\left.+e^{\text{i}\varphi_{3}}f_{\mathbf{r}}^{\dagger}f_{\mathbf{r}+\delta_{3}-\delta_{1}}-e^{-\text{i}\varphi_{3}}f_{\mathbf{r}+\delta_{2}}^{\dagger}f_{\mathbf{r}+\delta_{2}+\delta_{3}-\delta_{1}}+h.c.\right).
\end{aligned}
\end{equation}
In reciprocal space similarly,
\begin{equation}
\begin{aligned}
    H_{2}
    &=2\lambda\sum_{\mathbf{k}}\eta_{\mathbf{k}}^{\dagger}h^{(2)}(\mathbf{k})\eta_{\mathbf{k}}, \\
    h^{(2)}(\mathbf{k})
    &=\begin{bmatrix}
    h_{00}^{(2)}(\mathbf{k}) & h_{01}^{(2)}(\mathbf{k}) \\
    h_{10}^{(2)}(\mathbf{k}) & h_{11}^{(2)}(\mathbf{k})
    \end{bmatrix}
\end{aligned}
\end{equation}
with
\begin{equation}
    \begin{aligned}
    h_{00}^{(2)}(\mathbf{k})&=\cos(\mathbf{k}\cdot\delta_{2}+\mathbf{k}\cdot\delta_{3}+\varphi_{2}), \\
    h_{11}^{(2)}(\mathbf{k})&=\cos(\mathbf{k}\cdot\delta_{2}+\mathbf{k}\cdot\delta_{3}-\varphi_{2}), \\
    h_{01}^{(2)}(\mathbf{k})&=e^{\text{i}\varphi_{1}}\cos(\mathbf{k}\cdot\delta_{1}+\mathbf{k}\cdot\delta_{2})+\text{i}e^{\text{i}\varphi_{3}}\sin(\mathbf{k}\cdot\delta_{3}-\mathbf{k}\cdot\delta_{1}), \\
    h_{10}^{(2)}(\mathbf{k})&=h_{01}^{(2)*}(\mathbf{k}).
    \end{aligned}
\end{equation}
We may also assume real and equal amplitude $\rho$ for the 3rd NN couplings, thus
\begin{equation}
\begin{aligned}
H_{3}
&=\rho\sum_{\mathbf{r}}\left( e^{\text{i}\gamma_{1}}f_{\mathbf{r}}^{\dagger}f_{\mathbf{r}+2\delta_{1}}-e^{-\text{i}\gamma_{1}}f_{\mathbf{r}+\delta_{2}}^{\dagger}f_{\mathbf{r}+\delta_{2}+2\delta_{1}} \right. \\
&+e^{\text{i}\gamma_{2}}f_{\mathbf{r}}^{\dagger}f_{\mathbf{r}+2\delta_{2}}-e^{-\text{i}\gamma_{2} }f_{\mathbf{r}+\delta_{2}}^{\dagger}f_{\mathbf{r}+3\delta_{2}} \\
&\left.+e^{\text{i}\gamma_{3}}f_{\mathbf{r}}^{\dagger}f_{\mathbf{r}+2\delta_{3}}-e^{-\text{i}\gamma_{3}}f_{\mathbf{r}+\delta_{2}}^{\dagger}f_{\mathbf{r}+\delta_{2}+2\delta_{3}}+h.c. \right).
\end{aligned}
\end{equation}
In the reciprocal space we have
\begin{equation}
\begin{aligned}
    H_{3}
    &=2\rho\sum_{\mathbf{k}}\eta_{\mathbf{k}}^{\dagger}h^{(3)}(\mathbf{k})\eta_{\mathbf{k}}, \\
    h^{(3)}(\mathbf{k})
    &=\begin{bmatrix}
    h_{00}^{(3)}(\mathbf{k}) & 0 \\
    0 & h_{11}^{(3)}(\mathbf{k})
    \end{bmatrix}
\end{aligned}
\label{eq:ham3}
\end{equation}
with
\begin{equation}
    \begin{aligned}
    h_{00}^{(3)}(\mathbf{k})&=\cos(2\mathbf{k}\cdot\delta_{1}+\gamma_{1})+\cos(2\mathbf{k}\cdot\delta_{2}+\gamma_{2}) \\
    &+\cos(2\mathbf{k}\cdot\delta_{3}+\gamma_{3}), \\
    h_{11}^{(3)}(\mathbf{k})&=-\cos(2\mathbf{k}\cdot\delta_{1}-\gamma_{1})-\cos(2\mathbf{k}\cdot\delta_{2}-\gamma_{2}) \\
    &-\cos(2\mathbf{k}\cdot\delta_{3}-\gamma_{3}).
    \end{aligned}
\label{eq:ham3_entries}
\end{equation}
Finally, $H=J_{1}H_{1}+J_{2}H_{2}+J_{3}H_{3}$, where $J_{1, 2, 3}$ follows as same as the parameters in DMRG simulation. The 2nd and 3rd NN hoppings can open up a direct gap at each Dirac cone, leading to a Chern number $C=\pm1$ of the lower spinon band and the chiral edge states. Meanwhile the 3rd NN hoppings can break the degeneracy of two Dirac cones, giving rise to one particle-like spinon Fermi surface (SFS) around one Dirac point and a hole-like SFS around the other Dirac point. Due to single-occupancy constraint, the particle-like SFS and hole-like SFS are perfectly compensated at half filling.

\section{Central charge on the YC cylinders}

\begin{figure*}[t]
\includegraphics[width = 0.49\linewidth]{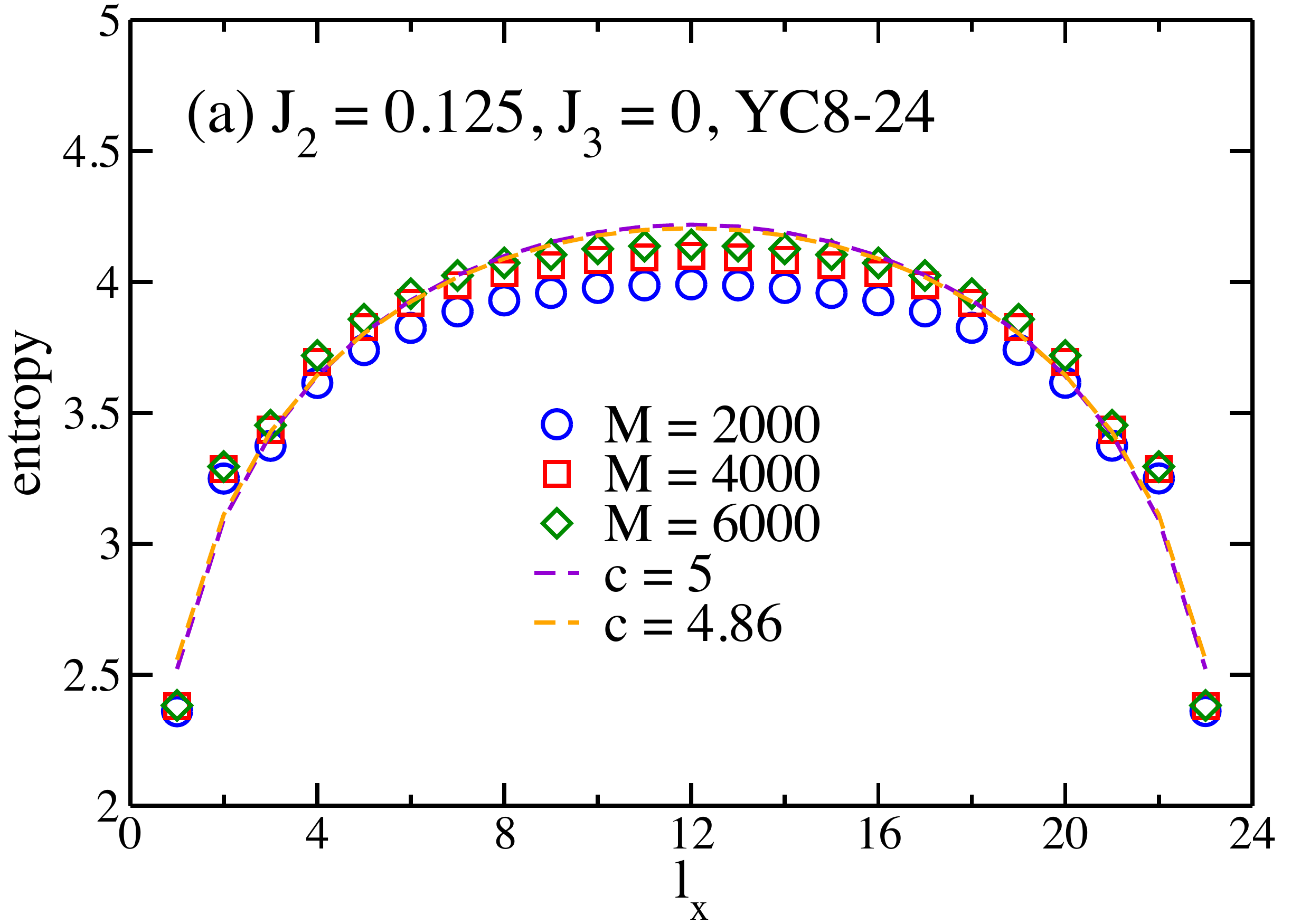}
\includegraphics[width = 0.49\linewidth]{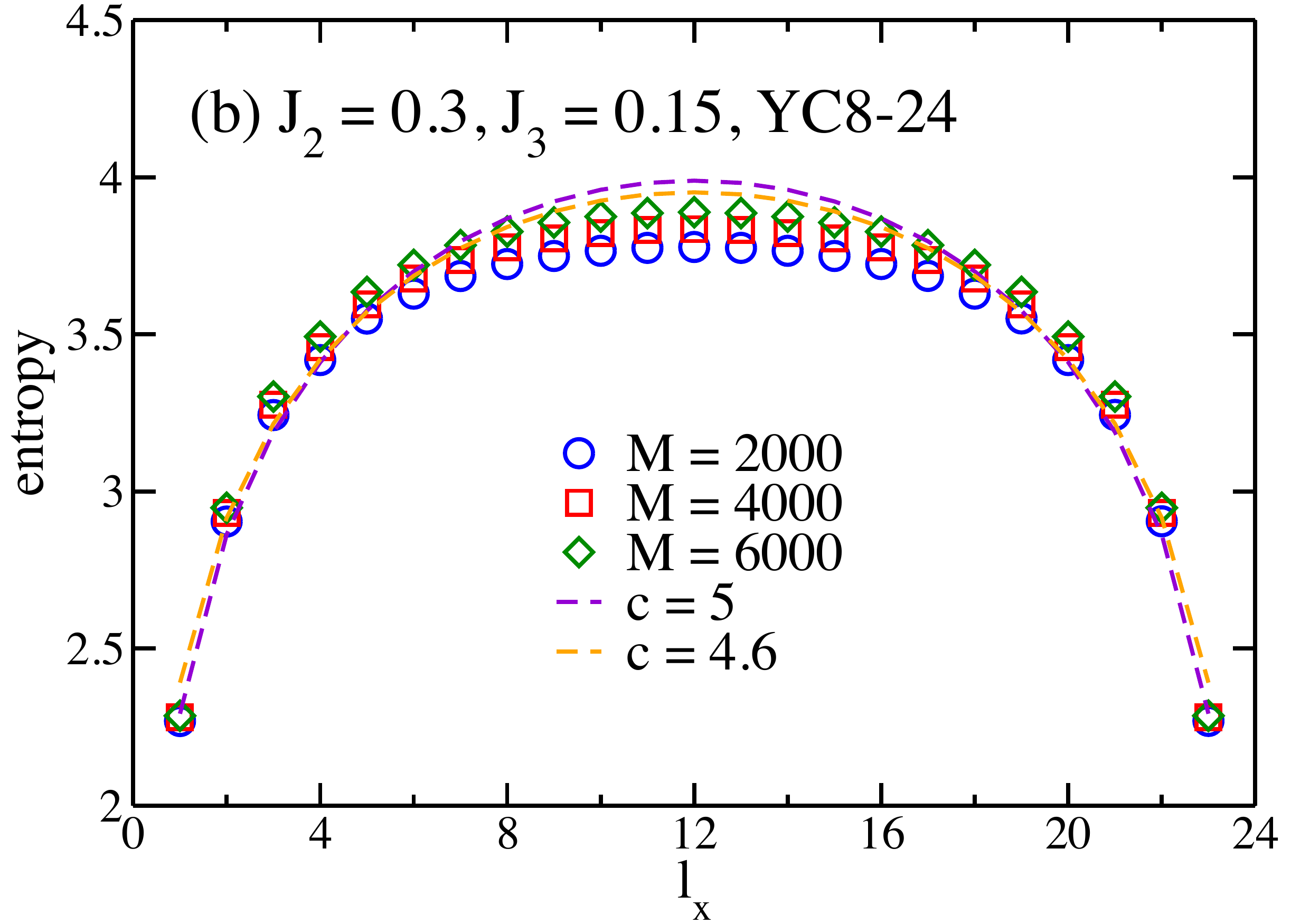}
\caption{Entanglement entropy versus subsystem length $l_x$ for (a) $J_2 = 0.125, J_3 = 0$ and (b) $J_2 = 0.3, J_3 = 0.15$ on the YC8-24 cylinder by keeping different $SU(2)$ state numbers $M$. The dashed lines denote the fitting of the entropy that are obtained by keeping $6000$ $SU(2)$ states by using the formula $S = (c/6) \ln( (L_x / \pi) \sin(l_x \pi / L_x)) + g$. The orange fittings are the best fittings without any assumption. The purple lines are the fittings with the fixed $c = 5.0$, which are close to the best fittings. Thus, the results are consistent with those on the YC8-16 cylinder.}
\label{figsup:entropy}
\end{figure*}

Below we discuss the central charge of the staggered flux state on YC cylinders, as illustrated in Fig.~4 (c)-(d) of the main text. Consider the 1D channels of quantized momenta (dashed orange lines in Fig.~4 of the main text) cross each fermi pockets $2n$ times for each spin species, so that in total there are $N_w=4n$ pairs of right and left movers among the gapless 1d modes. They can be labeled by their spin index $\sigma=\uparrow,\downarrow$, particle/hole pocket index $\mu=\pm1$, chirality index $\chi=\pm1$ for right and left movers, and flavor index $1\leq\rho\leq n$. Each gapless 1d mode of fermionic spinons on the YC cylinder can be bosonized as
\begin{equation}
\psi_{\chi,\mu,\rho,\sigma}({\bf r})\sim\kappa_{\chi,\mu,\rho,\sigma}e^{i\phi_{\chi,\mu,\rho,\sigma}(x_1)+i{\bf k}^{\chi,\mu,\rho}\cdot{\bf r}},~~~1\leq\rho\leq n=N_w/4.
\end{equation}
where $\kappa$ represents the Klein factors. Due to ``staggered'' magnetic translation symmetry (Eq.~(5) of the main text), the momenta of the 1D modes on particle and hole pockets are related by
\begin{equation}
{\bf k}^{\chi,+,\rho}\equiv {\bf k}^{\chi,\rho},~~~~~{\bf k}^{\chi,-,\rho}=\frac12{\bf b}_1-{\bf k}^{\chi,\rho}
\end{equation}
where ${\bf b}_{1,2}$ are reciprocal lattice vectors dual to Bravais lattice vectors ${\bf a}_{1,2}$. And the boson fields transform under translations as
\begin{eqnarray}\notag
&\phi_{\chi,\mu,\rho,\sigma}(x_1)\overset{T_1}\longrightarrow \phi_{\chi,\mu,\rho,\sigma}(x_1+a_1)+{\bf k}^{\chi,\mu,\rho}\cdot{\bf a}_1,\\
&\phi_{\chi,\mu,\rho,\sigma}(x_1)\overset{T_2}\longrightarrow-\phi_{\chi,-\mu,\rho,\sigma}(x_1)+{\bf k}^{\chi,\mu,\rho}\cdot{\bf a}_2.\label{1d YC:translation sym}
\end{eqnarray}
Due to the single-occupancy constraint of the spinons on each site, the total density fluctuation of the 1D gapless modes is forbidden:
\begin{equation}\label{density fluctuation}
  \delta\rho(x_1)=\frac1{2\pi}\partial_{x_1}(\sum_{\chi,\mu,\rho,\sigma}\chi\phi_{\chi,\mu,\rho,\sigma})\equiv0
\end{equation}
This is a consequence of the $U(1)$ gauge field coupled to the fermionic spinons. Therefore the low effective theory of the mean-field spinon Fermi surfaces coupled to a $U(1)$ gauge field, when placed on a finite YC cylinder, can be written as
\begin{equation}
\mathcal{L}_\text{YC}=\frac1{4\pi}\sum_{\chi,\mu,\rho,\sigma}\phi_{\chi,\mu,\rho,\sigma}(\chi\partial_t- v_{\chi,\rho}\partial_{x_1})\phi_{\chi,\mu,\rho,\sigma}-g_0\cos(\sum_{\chi,\mu,\rho,\sigma}\chi\phi_{\chi,\mu,\rho,\sigma})+\cdots
\end{equation}
where $v_{\chi,\rho}>0$ is the fermi velocity of each 1D mode $\phi_{\chi,\mu,\rho,\sigma}$. The cosine term $\propto g_0$ plays the role of $U(1)$ gauge field, removing the total density fluctuation Eq.~\eqref{density fluctuation} from the low-energy physics. Clearly the effective action preserves both translations $T_{1,2}$ in Eq.~\eqref{1d YC:translation sym} and $SU(2)$ spin rotational symmetry. 

Notice that the $U(1)$ gauge fluctuations (i.e. the $g_0$ term in effective action) reduces the total central charge $c$ of the YC cylinder from $N_w=4n$ to $N_w-1$, gapping out the total density mode. Also, we have not included any interaction between fermionic spinons in the above effective 1d action, which can further reduce the central charge to an integer smaller than $N_w-1$. Therefore we have shown that the central charge for the mean-field spinon Fermi surfaces in Fig.~4 (c)-(d) of the main text, is upper bounded by $c \leq N_w-1 = 4n-1$. This means $c \leq 3$ on YC6 cylinder ($n=1,N_w=4$) in Fig.~4(c), and $c\leq7$ on YC8 cylinder ($n=2,N_w=8$) in Fig.~4(d) of the main text.

\bibliography{triangle}

\end{document}